\documentclass[10pt,journal]{IEEEtran}
\usepackage{graphicx}
\usepackage{amssymb,amsmath,mathrsfs}
\begin{document}
\title{Waterfilling Theorems for Linear Time-Varying Channels and Related Nonstationary Sources}

\author{Edwin~Hammerich,~\IEEEmembership{Member,~IEEE}
\thanks{The material in this paper was presented in part at the 2014 IEEE International Symposium 
on Information Theory \cite{Ham2014}.

The author is with the Ministry of Defence, Kulmbacher Str.\ 58--60, D-95030 Hof, 
Germany (e-mail: edwin.hammerich@ieee.org).}}
\maketitle

\begin{abstract}
The capacity of the linear time-varying (LTV) channel, a continuous-time LTV filter with additive white Gaussian 
noise, is characterized by waterfilling in the time--frequency plane. 
Similarly, the rate distortion 
function for a related nonstationary source is characterized by reverse waterfilling in the 
time--frequency plane. Constraints on the average energy or on the squared-error 
distortion, respectively, are used. The source is formed by the white Gaussian noise response of 
the same LTV filter as before. The proofs of both waterfilling theorems rely on a Szeg\H{o} 
theorem for a class of operators associated with the filter. A self-contained proof 
of the Szeg\H{o} theorem is given. The waterfilling theorems compare well with the classical results of 
Gallager and Berger. In the case of a nonstationary source, it is observed that the part of the classical 
power spectral density is taken by the Wigner--Ville spectrum. The present approach is based on the spread 
Weyl symbol of the LTV filter, and is asymptotic in nature. For the spreading factor, a lower bound is 
suggested by means of an uncertainty inequality.
\end{abstract}
\begin{IEEEkeywords}
Channel capacity, linear time-varying (LTV) channel, nonstationary source, rate distortion function, 
Szego theorem, time--frequency transfer function, uncertainty.
\end{IEEEkeywords}

\newtheorem{definition}{Definition}
\newtheorem{theorem}{Theorem}
\newtheorem{lemma}{Lemma}
\newtheorem{remark}{Remark}
\newtheorem{example}{Example}

\renewcommand{\d}{\mathrm{d}}
\newcommand{\e}{\mathrm{e}}
\renewcommand{\i}{\mathrm{i}}
\renewcommand{\P}{\boldsymbol{P}}
\newcommand{\Q}{\boldsymbol{Q}}
\newcommand{\A}{\boldsymbol{A}}
\newcommand{\Atilde}{\boldsymbol{\tilde{A}}}
\newcommand{\Ahat}{\boldsymbol{\hat{A}}}
\newcommand{\B}{\boldsymbol{B}}
\newcommand{\x}{\boldsymbol{x}}
\newcommand{\y}{\boldsymbol{y}}
\newcommand{\z}{\boldsymbol{z}}
\newcommand{\bsalpha}{\boldsymbol{\alpha}}
\newcommand{\bsbeta}{\boldsymbol{\beta}}
\renewcommand{\S}{\mathscr{S}}
\newcommand{\tr}{\mathrm{tr}}
\newcommand{\PrStarPr}{\P_r\!\!^*\P_r}
\newcommand{\LambertW}{\mathrm{W}}
\newcommand{\SNR}{\mathrm{SNR}}
\newcommand{\SDR}{\mathrm{SDR}}

\section{Introduction}
\IEEEPARstart{T}{he} characterization of the capacity of continuous-time channels with an average power 
constraint by waterfilling in the frequency domain, going back to Shannon \cite{Shannon1949}, has been given 
by Gallager \cite{Gallager} 
for linear time-invariant (LTI) channels in great generality. At least since the advent of 
mobile communications, there has been a vivid interest in similar results for LTV channels; see 
\cite{Barbarossa}, \cite{Jung}, \cite{Farrell}, \cite{DSBS} to cite only a few. Although most wireless 
communication channels are modeled by \emph{random} LTV filters \cite{Bello}, \cite{DSBS}, a waterfilling 
characterization of the capacity of deterministic LTV channels might also be of interest. Furthermore, many 
nonstationary continuous-time sources can be described as the response of an LTV filter to white Gaussian 
noise. It is therefore natural to ask for a solution to the dual problem, namely the reverse waterfilling 
characterization of the rate distortion function for such sources with a fidelity criterion. The 
classical answer to this question in the case of a stationary source, already outlined by Kolmogorov in 
\cite{Kolm}, has been given by Berger \cite{Berger} for a broad class of stationary random processes. 
Since then, until quite recently \cite{KiGo}, no similar results for nonstationary sources have been 
reported. Within the framework of time--frequency analysis, treating the time--frequency 
plane ``as a whole" \cite{Daub}, we present waterfilling solutions to both problems (with constraints on 
the average energy in the case of the channel and on the squared-error distortion in the case of the source).

We consider integral operators $\P$ from the Hilbert space 
$L^2(\mathbb{R})$ of square-integrable functions $f:\mathbb{R}\rightarrow\mathbb{C}\cup\{\infty\}$ 
into itself of the form
\begin{equation}
  (\P f)(t)=\int_{-\infty}^{\infty}h(t,t')f(t')\,\d t' \label{Op_P}
\end{equation}
with the kernel $h\in L^2(\mathbb{R}^2)$, i.e., Hilbert--Schmidt (HS) operators on $L^2(\mathbb{R})$ \cite{Reed}. 
Every such operator has a unique Weyl symbol $p=\sigma_{\P }\in L^2(\mathbb{R}^2)$ so that Eq.~\eqref{Op_P} 
may be written as \cite{Pool}, \cite{KoHl}
\begin{equation}
  (\P f)(t)=\frac{1}{2\pi}\iint_{\mathbb{R}^2}
                 p\left(\frac{t+t'}{2},\omega\right)\e^{\i(t-t')\omega}f(t')\,\d t'\,\d\omega. \label{Eq_pP}
\end{equation}
The Weyl symbol, a concept originating in quantum mechanics \cite{Gosson11}, \cite{Folland}, \cite{Groch}, is now a standard tool 
for the description of LTV systems \cite{MaHl} (because of its physical provenance, we shall often 
switch between variables 
$t,\omega$ and $x,\xi$ standing for time, angular frequency and the corresponding phase space coordinates). The operator 
\eqref{Op_P}, regarded as an LTV filter for finite-energy signals $f(t)$, will play a central role in our investigations. However, for 
the formulation of problems it 
will be necessary to replace $\P$ with the operator $\P_r:L^2(\mathbb{R})\rightarrow L^2(\mathbb{R})$ having the 
\emph{spread} Weyl symbol $\sigma_{\P_r}(t,\omega)=p_r(t,\omega)\triangleq p(t/r,\omega/r)$, where $r\ge1$ is 
the spreading factor. Eq.~\eqref{Op_P} then turns into
\begin{equation}
  (\P_r f)(t) = \int_{-\infty}^\infty h(r,t,t')f(t')\,\d t',  \label{Op_Pr}
\end{equation}
where $h(r,\cdot,\cdot)\in L^2(\mathbb{R}^2)$ denotes the kernel, now depending on $r$. It is not difficult to express 
$h(r,t,t')$ in terms of $h(t,t')$ and $r$; however, we shall rarely make use of that representation since 
the Weyl symbol appears to be the appropriate filter description in our context. Although other choices are possible for 
that symbol (also called the time--frequency transfer function; see \cite{MaHl} for a systematic overview), the Weyl symbol 
excels due to some unique properties, one of them being most helpful later on. There is one other choice for the description 
of LTV filters: the spreading function \cite{Bello}, \cite{Groch}, \cite{MaHl}. This is the two-dimensional (symplectic) 
Fourier transform of, in our case, the Weyl symbol $\sigma_{\P}$,
\[
  \hat{\sigma}_{\P}(\tau,\nu)=\frac{1}{2\pi}\iint_{\mathbb{R}^2}\e^{-\i(x\nu-\tau\xi)}\sigma_{\P}(x,\xi)\,\d x\,\d\xi,
\]
and its popularity in mobile communications comes from the fact that the representation
\[
  (\P f)(t)=\frac{1}{2\pi}\iint_{\mathbb{R}^2}\hat{\sigma}_{\P}(\tau,\nu)\e^{-\i\tau\nu/2}f(t-\tau)\e^{\i t\nu}\,\d\tau\,\d\nu
\]
allows a simple interpretation of the operator in terms of a weighted superposition of time delays $\tau$ and Doppler 
shifts $\nu$ of the input signal. Because of 
$\hat{\sigma}_{\P_r}(\tau,\nu)=r^2\hat{\sigma}_{\P}(r\tau,r\nu)$ we observe increasing concentration of the 
spreading function $\hat{\sigma}_{\P_r}$ of operator $\P_r$ around the origin of the $\tau,\nu$-plane as $r\rightarrow\infty$. 
This behaviour, shared by many practical LTV filters and termed \emph{underspread} in 
\cite{Kozek97}, \cite{MaHl}, is therefore also peculiar to our setting (where, in principle, $r$ tends 
to infinity). However, it remains to be remarked
that the spreading function would not be the proper means for formulating the subsequent 
waterfilling theorems, Theorem~\ref{WFT1} and Theorem~\ref{WFT2}.

The present paper evolves from previous work presented in \cite{Ham2014}. We now give a brief overview 
of the contributions of our paper with emphasis on extensions and modifications compared to 
\cite{Ham2014}; for details, refer to the text. The LTV filters, initially arbitrary HS operators, are 
later restricted to those having Weyl symbols in the Schwartz space of rapidly decreasing 
functions (thus including the bivariate Gaussian function used in \cite{Ham2014}). The waterfilling theorem for the 
capacity of the LTV channel is now stated in terms of the 
reciprocal squared modulus of the spread Weyl symbol of the LTV filter. Similarly, the reverse 
waterfilling theorem for the rate distortion function for the nonstationary source is stated in terms of the 
squared modulus of the spread Weyl symbol of the LTV filter. A major difference from 
\cite{Ham2014} is the statement of a new Szeg\H{o} theorem, which is now general enough to cover a 
large class of operators. 
For part of the proof of the Szeg\H{o} theorem we resort to a powerful asymptotic expansion having 
its roots in semiclassical physics \cite{Gosson11}, \cite{Folland}, \cite{Robert}. Since our results are asymptotic in 
nature, there is a need to give a lower bound for the spreading factor so that the formulas 
in the waterfilling theorems yield useful approximations. A lower bound is suggested by means of the Robertson--Schr\"{o}dinger 
uncertainty inequality \cite{Gosson11}. Several concrete examples will illustrate our results.

\section{Mathematical Preliminaries} \label{Sec_MathPrel}
In the present section, we fix the notation and compile some mathematical concepts and results associated 
with the LTV filter \eqref{Op_Pr}. In Section~\ref{Sec_Fundamentals}, it will be sufficient to restrict 
ourselves to the spreading factor $r=1$, therefore it is omitted; generalizations to the case $r\ge1$, 
mostly obvious, will be addressed as needed in the subsequent sections.

\subsection{Notation}
The following notations will be adopted: 
The inner product in $L^2(\mathbb{R})$ is denoted by  
$\langle f_1,f_2\rangle=\int_{-\infty}^\infty f_1(x)\overline{f_2(x)}\,\d x$, and 
$\|f\|=\langle f,f\rangle^{1/2}$ is the corresponding norm. For an operator 
$\A:L^2(\mathbb{R})\rightarrow L^2(\mathbb{R})$, its adjoint  
$\A^*:L^2(\mathbb{R})\rightarrow L^2(\mathbb{R})$ is defined by the condition 
$\langle\A f_1,f_2\rangle=\langle f_1,\A ^*f_2\rangle\,\forall f_1,f_2\in L^2(\mathbb{R})$; $\A$ is called 
self-adjoint if $\A^*=\A$. 
$\S(\mathbb{R}^n),\,n=1,2,$ is the Schwartz space of rapidly decreasing functions on 
$\mathbb{R}^n$ (cf. \cite{Groch}); if $n=2$ and the function $u$ additionally depends on the 
parameter $r$, $u=u(r,x,\xi)$, then $u\in\S(\cdot,\mathbb{R}^2)$ means
\[
  \sup_{x,\xi}|x^{\beta_1}\xi^{\beta_2}\partial_x^{\alpha_1}\partial_\xi^{\alpha_2}u(r,x,\xi)|\le 
                                                                        C_{\bsalpha\bsbeta}<\infty
\]
for all $\bsalpha=(\alpha_1,\alpha_2),\bsbeta=(\beta_1,\beta_2)\in\mathbb{N}_0^2$, where the constants 
$C_{\bsalpha\bsbeta}$ do not depend on $r$. $L^2_{\mathbb{R}}(\mathbb{R})$ is the real Hilbert space of 
real-valued functions in $L^2(\mathbb{R})$.
\subsection{Fundamental Concepts and Results} \label{Sec_Fundamentals}
\subsubsection{Weyl correspondence}
The Weyl symbol $\sigma_{\P}$ of the HS operator $\P$ in \eqref{Op_P} is given by the equation 
(sometimes called the Wigner transform) \cite{Pool}, \cite{KoHl}
\begin{equation}
  \sigma_{\P }(x,\xi)=\int_{-\infty}^\infty \e^{-\i\xi x'}h\left(x+\frac{x'}{2},
                                                              x-\frac{x'}{2}\right)\,\d x'. \label{WT}
\end{equation}
The linear mapping $\P\mapsto p=\sigma_{\P}$ defined by \eqref{WT} establishes a one-to-one correspondence 
between all HS operators on $L^2(\mathbb{R})$ and all functions $p\in L^2(\mathbb{R}^2)$ \cite{Pool}, 
\cite{Groch}. Moreover, it holds (here and hereafter, double integrals extend over $\mathbb{R}^2$)
\begin{equation}
  \frac{1}{2\pi}\iint|p(x,\xi)|^2\,\d x\,\d\xi=\iint|h(x,y)|^2\,\d x\,\d y.             \label{JCTP}
\end{equation}
The above mapping (or rather its inverse) is called Weyl correspondence \cite{Groch}.

\subsubsection{Singular value decomposition (SVD)} \label{Sec_SVD}
Every HS operator $\P$ on $L^2(\mathbb{R})$ is compact and so is its adjoint $\P^*$ \cite{Reed}. 
Define the self-adjoint operator $\A \triangleq\P^*\P$ on $L^2(\mathbb{R})$. $\A$ is positive because 
$\langle \A f,f\rangle=\langle \P f,\P f\rangle\ge0\,\forall f\in L^2(\mathbb{R})$, 
and compact since one factor, say, $\P$, is compact. Therefore, $\P$ has the SVD \cite{Reed}, \cite[Th.~8.4.1]{Gallager}
\begin{equation}
  (\P f)(x)=\sum_{k=0}^N \sqrt{\lambda_k}\,\langle f,f_k\rangle g_k(x),  \label{SVD}
\end{equation}
where $\{f_0,\ldots,f_N\}$, $\{g_0,\ldots,g_N\}$ ($N\in\mathbb{N}_0$ or $N=\infty$) form orthonormal systems in 
$L^2(\mathbb{R})$, and $\lambda_0\ge\lambda_1\ge\ldots>0$ are the non-zero eigenvalues of $\A$ (counting 
multiplicity) with the corresponding eigenfunctions $f_k$; the functions $g_k$ are defined by 
$g_k=\P f_k/\sqrt{\lambda_k}$, the positive numbers $\sqrt{\lambda_k},\,k=0,\ldots,N$, being the non-zero 
\emph{singular values} of $\P$. If $\P$ maps $L^2_{\mathbb{R}}(\mathbb{R})$ into itself, then the 
functions $f_k,g_k$ will be real-valued. Without loss of generality (w.l.o.g.) we shall assume that $N=\infty$ (otherwise, 
put $\lambda_k=0$ and choose $f_k,g_k$ anyway for $k>N$). Then always $\lambda_k\rightarrow0$ as $k\rightarrow\infty$.

\subsubsection{Traces of operators}
By Eq.~\eqref{SVD}, the kernel of operator $\P$ in \eqref{Op_P} has the form 
$h(x,y)=\sum_{k=0}^\infty \sqrt{\lambda_k}g_k(x)\overline{f_k(y)}$ from 
where we readily obtain $\iint |h(x,y)|^2\,\d x\,\d y=\sum_{k=0}^\infty \lambda_k$. In combination with 
\eqref{JCTP}, this results in the useful equation
\begin{equation}
  \tr\,\A\triangleq\sum_{k=0}^\infty \lambda_k
                                  =\frac{1}{2\pi}\iint|p(x,\xi)|^2\,\d x\,\d\xi<\infty. \label{ID1}
\end{equation}
Since $\tr\,\A$ (the trace of $\A$) is finite, $\A$ is of \emph{trace class} (see \cite{Reed} for a 
general definition of trace class operators).

In Section \ref{Sec_VI}, the operator $\Atilde\triangleq\P\P^*$ will be considered. Plugging 
$\P^*f\in L^2(\mathbb{R})$ for $f\in L^2(\mathbb{R})$ in \eqref{SVD} we get for $\Atilde$ the  
representation $(\Atilde f)(x)=\int K_{\Atilde}(x,y)f(y)\,\d y$ with the kernel
\begin{equation}
  K_{\Atilde}(x,y)=\sum_{k=0}^\infty\lambda_k g_k(x)\overline{g_k(y)}.   \label{K_PPstar}
\end{equation}
$\Atilde$ has the same eigenvalues as $\A$. Furthermore, since we are dealing with the \emph{Weyl} 
symbol we have the simple rule
\begin{equation}
  \sigma_{\P ^*}(x,\xi)=\overline{\sigma_{\P}(x,\xi)}. \label{Eq_sigmabar}
\end{equation}
Hence, Eq.~\eqref{ID1} holds by analogy for operator $\Atilde$ (just replace ``$\A$" with 
``$\Atilde$").

In quantum mechanics, an operator on $L^2(\mathbb{R})$ is called a density operator, if it is 
1) self-adjoint, 2) positive and 3) of trace class with trace one \cite{Gosson11}. 
Apparently, the above operators $\A,\Atilde$ enjoy all these properties, with the exception of the 
very last. We give them a name:
\begin{definition} \label{Def_QDO}
A quasi density operator (QDO) is an operator on $L^2(\mathbb{R})$ of the form $\P^*\!\P$ or $\P\P^*$, 
where $\P:L^2(\mathbb{R})\rightarrow L^2(\mathbb{R})$ is an HS operator.
\end{definition}
\begin{remark} In \cite{Gosson08} it is noted that \emph{any} self-adjoint, positive operator on 
$L^2(\mathbb{R})$ of trace class allows factorizations as given in Def.~\ref{Def_QDO}; the 
above narrow-sense meaning of QDO will be sufficient for our purposes. 
\end{remark}

The following result is key to our paper: If the operator 
$\B:L^2(\mathbb{R})\rightarrow L^2(\mathbb{R})$ has a Weyl symbol 
$\sigma_{\B}\in\S(\mathbb{R}^2)$, then $\B$ is of trace class and its trace is given by the \emph{trace 
rule} \cite{Janssen}
\begin{equation}
  \tr\,\B=\frac{1}{2\pi}\iint\sigma_{\B}(x,\xi)\,\d x\,\d\xi. \label{Eq_tracerule}
\end{equation}
Refer to \cite{Gosson11} concerning the smoothness assumption and for a proof.

\subsubsection{Bound on eigenvalues} \label{Sec_CV}
If the function $a=a(x,\xi):\mathbb{R}^2\rightarrow\mathbb{C}$ is differentiable up to the sixth order 
and it holds
\begin{equation}
  \sup_{x,\xi}|\partial_x^{\alpha_1}\partial_\xi^{\alpha_2}a(x,\xi)|\le C_{\bsalpha}<\infty \label{Ineq_CV}
\end{equation}
for all $\bsalpha=(\alpha_1,\alpha_2)\in I=\{0,1,2,3\}^2$, then the operator $\A$ defined 
by the Weyl symbol $a$ is a \emph{bounded} operator from $L^2(\mathbb{R})$ into itself, and it holds 
\[
  \|\A f\|\le c_0C\,\|f\|,\,f\in L^2(\mathbb{R}),
\]
where $C=\sum_{\bsalpha\in I}C_{\bsalpha}$ and $c_0$ is a certain constant 
not depending on the operator. This is the famous theorem of Calder\'{o}n--Vaillancourt \cite{Calderon}, 
\cite{Folland}. Consequently, the absolute value $|\lambda|$ of every eigenvalue $\lambda$ of 
$\A$ is bounded by $c_0C$.

\section{Channel Model and Discretization} \label{Sec_III}
\begin{figure}
\setlength{\unitlength}{1cm}
\begin{picture}(8.5,4)
  \put(3,1){\framebox(2,1){$p_r(t,\omega)$}}
  \put(2,1.5){\vector(1,0){1}}
  \put(1.0,1.4){$f(t)$}
  \put(6.25,1.5){\vector(1,0){1.0}}
  \put(5.875,1.375){\makebox(0.25,0.25){$+$}}
  \put(5,1.5){\vector(1,0){0.75}}
  \put(7.5,1.45){$\tilde{g}(t)$}
  \put(6,3){\vector(0,-1){1.25}}
  \put(5.7,3.2){$n(t)$}
  \put(6,1.5){\circle{0.5}}
  \put(6.5,3.2){\parbox{1.75cm}{\footnotesize white Gaussian noise with PSD $N_0/2=\theta^2$}}
  \put(0.6,0.5){\parbox{1.75cm}{\footnotesize finite-energy, real-valued}}
  \put(3.4,0.5){\parbox{2cm}{\footnotesize LTV filter}}
\end{picture}
\caption{Model of the LTV channel. The Weyl symbol $p_r(t,\omega)$ acts as a time--frequency 
transfer function; $r\ge1$ is the spreading factor.} \label{Fig_1}
\end{figure}
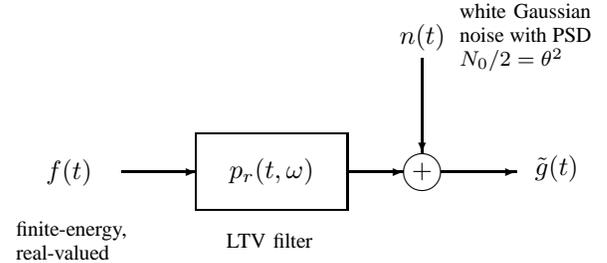

We consider for any spreading factor $r\ge 1$ held constant the LTV channel
\begin{equation}
   \tilde{g}(t)=(\P _rf)(t)+n(t),\,-\infty<t<\infty,  \label{LTV_Ch}
\end{equation}
where $\P _r$ is the LTV filter \eqref{Op_Pr}, the real-valued filter input signals $f(t)$ are of finite 
energy and the noise signals  $n(t)$ at the filter output are realizations of white Gaussian noise with 
two-sided power spectral density (PSD) $N_0/2=\theta^2>0$. Moreover, we assume throughout that 
the kernel $h(t,t')$ of operator $\P$ in \eqref{Op_P} is real-valued; observe that due to
\begin{multline*}
 h(r,t,t')
  =rh((r^{-1}(t+t')+r(t-t'))/2,\\(r^{-1}(t+t')-r(t-t'))/2),
\end{multline*}
then also the kernel $h(r,t,t')$ of operator $\P_r$ will be real-valued so that $\P_r$ maps 
$L^2_{\mathbb{R}}(\mathbb{R})$ into itself. This channel is depicted in Fig.~\ref{Fig_1}.  

We now reduce the LTV channel \eqref{LTV_Ch} to a (discrete) vector Gaussian channel, 
following the approach in \cite{Gallager} for LTI channels; our analysis is greatly simplified by 
the restriction to finite-energy input signals. For the SVD of operator $\P_r$ the $r$-dependent 
operator $\A(r)\triangleq\PrStarPr$ has to be considered; since eigenvalues 
$\lambda_k$ and (eigen-)functions $f_k,\,g_k$ in the SVD now also depend on $r$, this will be 
indicated by a superscript~$\cdot\,^{(r)}$. Then, by Eq.~\eqref{SVD}, the LTV filter \eqref{Op_Pr} has the 
SVD
\begin{equation}
  (\P_rf)(t)=\sum_{k=0}^\infty [\lambda_k^{(r)}]^{\frac{1}{2}}a_k\,g_k^{(r)}\!(t),          \label{SVD_r}
\end{equation}
where the coefficients are $a_k=\langle f,f_k^{(r)}\rangle,\,k=0,1,\ldots,$ and $\{g_k^{(r)};k=0,1,\ldots\}$ 
forms an orthonormal system in $L^2(\mathbb{R})$. Recall from Section~\ref{Sec_SVD} that the functions 
$f_k^{(r)}\!,\,g_k^{(r)}$ are real-valued. The perturbed filter 
output signal $g=\P_rf$, $\tilde{g}(t)=g(t)+n(t)$, is passed through a bank of matched 
filters with impulse responses $h_k(t)=g_k^{(r)}(-t),\,k=0,1,\ldots\,.$ The matched filter output 
signals are sampled at time zero to yield $\langle\tilde{g}(t),h_k(-t)\rangle=b_k+n_k$, where 
$b_k=\langle g(t),h_k(-t)\rangle=[\lambda_k^{(r)}]^{1/2}a_k$, and the detection errors 
$n_k=\langle n(t),h_k(-t)\rangle$ are realizations of independent identically distributed (i.i.d.)
zero-mean Gaussian random variables $N_k$ with the variance $\theta^2$, $N_k\sim\mathcal{N}(0,\theta^2)$. 
From the detected values $\hat{b}_k=b_k+n_k$ we get the estimates 
$\hat{a}_k=[\lambda_k^{(r)}]^{-1/2}\hat{b}_k=a_k+z_k$ for the coefficients $a_k$ of the input 
signal $f$, where $z_k$ are realizations of independent Gaussian random variables 
$Z_k\sim\mathcal{N}(0,\theta^2/\lambda_k^{(r)})$. Thus, we are led to the infinite-dimensional 
vector Gaussian channel
\begin{equation}
  Y_k=X_k+Z_k,\,k=0,1,\ldots,  \label{LTV_discr}
\end{equation}
where the noise $Z_k$ is distributed as described. Note that the noise PSD $\theta^2$, measured in 
watts/Hz, also has the physical dimension of an energy.

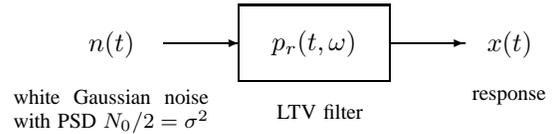
\begin{figure}
\setlength{\unitlength}{1cm}
\begin{picture}(8.5,3)
  \put(3.5,1){\framebox(2,1){$p_r(t,\omega)$}}
  \put(2.5,1.5){\vector(1,0){1}}
  \put(1.5,1.4){$n(t)$}
  \put(5.5,1.5){\vector(1,0){1.0}}
  \put(6.8,1.4){$x(t)$}
  \put(6.6,0.75){\footnotesize response}
  \put(0.5,0.5){\parbox{2.6cm}{\footnotesize white Gaussian noise with PSD $N_0/2=\sigma^2$}}
  \put(4,0.5){\parbox{2cm}{\footnotesize LTV filter}}
\end{picture}
\caption{Model of the nonstationary source} \label{Fig_2}
\end{figure}


\section{A Szeg\H{o} Theorem for Quasi Density Operators} \label{Sec_IV}
From now on to the end of the paper, we assume that the Weyl symbol $p$ of the HS operator $\P$ in 
\eqref{Eq_pP} is in the Schwartz space of rapidly decreasing functions, $p\in\S(\mathbb{R}^2)$.

Consider the QDO $\A=\P^*\!\P$ and generalize it as above to the operator $\A(r)=\PrStarPr$, $r\ge1$ (being again a QDO). We now 
state and prove a Szeg\H{o} theorem for $\A(r)$. Szeg\H{o} theorems like the subsequent 
Theorem~\ref{SzegoTh} are not new \cite{Widom}, 
\cite{Janssen}, \cite{Fei}, \cite{Oldfield}, but all the Szeg\H{o} theorems we are aware of are inadequate for our purposes. 
The proof of Lemma~\ref{Lemma_1} (see below) rests on an asymptotic expansion of the $n$th 
power of $\A(r)$. Asymptotic expansions such as that (there are different kinds of estimating the error!) have a long tradition 
in semiclassical physics and the theory of pseudodifferential operators \cite{Widom}, \cite{Folland}; 
rigorous proofs, however, are sometimes hard to find. A complete proof of the following Lemma~\ref{power_lemma}, which is 
perhaps closest to results of \cite{Robert}, is shifted to the Appendix. Although we need the lemma 
only in the case of $m=1$, it would not be natural to omit a full statement of it:
\begin{lemma}\label{power_lemma} For any $n\in\mathbb{N}$, the Weyl symbol of the operator 
$\A^n(r)\triangleq[\A(r)]^n,\,r\ge1,$ has the asymptotic expansion
\begin{equation}
  \sigma_{\A^n(r)}(x,\xi)\sim\sum_{k=0}^\infty r^{-2k}a_k(x/r,\xi/r),  \label{AE}
\end{equation}
where $a_0(x,\xi)=|p(x,\xi)|^{2n}$, $a_k\in\S(\mathbb{R}^2)$ else, and Eq.~\eqref{AE} means that 
for all $m\in\mathbb{N}$ it holds
\begin{multline}
  \sigma_{\A^n(r)}(x,\xi)=\sum_{k=0}^{m-1} r^{-2k}a_k(x/r,\xi/r)\\
                                                   +r^{-2m}R_m(r,x/r,\xi/r),\nonumber
\end{multline}
where $R_m=R_m(r,x,\xi)\in\S(\cdot,\mathbb{R}^2)$.
\end{lemma}
\begin{IEEEproof} See Appendix. \end{IEEEproof}
Asymptotically, i.e., as $r\rightarrow\infty$, $a_0(x/r,\xi/r)$ is the dominant part of the asymptotic 
expansion \eqref{AE}. As customary in the theory of pseudodifferential operators (cf., e.g., \cite{Folland}, 
\cite{Janssen}), the expression $|p_r(x,\xi)|^{2n}$ will be called the \textit{principal symbol} of operator $\A^n(r)$. 
Observe that the Weyl symbol of the $n$th power of the operator $\Atilde(r)=\P_r\P_r^*,r\ge1,$ has an asymptotic 
expansion analogous to that of $\A^n(r)$ and the principal symbols of both operators are identical.

\begin{definition} \label{def_2} For any two functions $A,\,B:[1,\infty)\rightarrow\mathbb{R}$ 
the notation $A\doteq B$ means
\[
  \lim_{x\rightarrow\infty}\frac{A(x)-B(x)}{x^2}=0,
\]
or, equivalently, $A(x)=B(x)+o(x^2)$ as $x\rightarrow\infty$, where $o(\cdot)$ denotes the 
standard Landau little-o symbol.
\end{definition}
In our context, $x$ will always be the spreading factor $r\ge1$. Thus $A\doteq B$ implies that 
$A(r)/r^2=B(r)/r^2+\epsilon$ where $\epsilon\rightarrow 0$ as $r\rightarrow\infty$.

\begin{lemma}\label{Lemma_1} For any polynomial $G_N(x,z)=\sum_{n=1}^N c_n(x) z^n$ 
with bounded variable coefficients $c_n(x)\in\mathbb{R},\,x\ge 1,$ it holds
\[
  \sum_{k=0}^\infty G_N(r,\lambda_k^{(r)})\doteq\frac{1}{2\pi}
                       \iint_{\mathbb{R}^2}G_N(r,|p_r(x,\xi)|^2)\,\d x\,\d\xi.
\]
\end{lemma}
\begin{IEEEproof}
First, application of operator $\P_r^*$ to both sides of Eq.~\eqref{SVD_r} yields
\[
  \A(r)f=\sum_{k=0}^\infty\lambda_k^{(r)}\langle f,f_k^{(r)}\rangle f_k^{(r)}.
\]
So we get for any $f\in L^2(\mathbb{R})$ the expansion
\[
  G_N(r,\A(r))f=\sum_{k=0}^\infty G_N(r,\lambda_k^{(r)})\langle f,f_k^{(r)}\rangle f_k^{(r)}.
\]
Hence, operator $\boldsymbol{B}(r)\triangleq G_N(r,\A(r))$ is of trace class with the trace
\begin{equation}
  \tr\,\boldsymbol{B}(r)=\sum_{k=0}^\infty G_N(r,\lambda_k^{(r)}),   \label{trace_1}
\end{equation}
the series being absolutely converging since $G_N(x,0)=0\,\forall x\in[1,\infty)$.

Second, we use the trace rule \eqref{Eq_tracerule} to obtain
\begin{equation} \label{Eq_trBrInt}
  \tr\,\B(r)=\frac{1}{2\pi}\iint\sigma_{\B(r)}(x,\xi)\,\d x\,\d\xi,
\end{equation}
where $\sigma_{\boldsymbol{B}(r)}(x,\xi)$ is the Weyl symbol of operator $\boldsymbol{B}(r)$. By 
linearity of the Weyl correspondence, $\sigma_{\B(r)}(x,\xi)$ has the expansion 
\begin{equation} \label{Eq_sigmaBr}
   \sigma_{\B(r)}(x,\xi)=\sum_{n=1}^N c_n(r)\sigma_{\A^n(r)}(x,\xi).
\end{equation}
From Lemma~\ref{power_lemma}, taking $m=1$, we infer that
\[
  \iint\sigma_{\A^n(r)}(x,\xi)\,\d x\,\d\xi 
                         \doteq\iint|p_r(x,\xi)|^{2n}\,\d x\,\d\xi.
\]
Plugging \eqref{Eq_sigmaBr} into \eqref{Eq_trBrInt}, we obtain by means of the latter equation
\begin{align}
  \tr\,\boldsymbol{B}(r)&=\frac{1}{2\pi}\sum_{n=1}^N c_n(r)
                                   \iint\sigma_{\A^n(r)}(x,\xi)\,\d x\,\d\xi \nonumber\\
    &\doteq\frac{1}{2\pi}\sum_{n=1}^N c_n(r)
                          \iint|p_r(x,\xi)|^{2n}\,\d x\,\d\xi\nonumber\\
    &=\frac{1}{2\pi}\iint G_N(r,|p_r(x,\xi)|^2)\,\d x\,\d\xi.     \label{trace_2}
\end{align}
Eq.~\eqref{trace_2} in combination with Eq.~\eqref{trace_1} concludes the proof.
\end{IEEEproof}

Lemma~\ref{power_lemma} shows that in the case of $n=1$ and, say, $m=1$, the Weyl symbol 
$a(r,x,\xi)=\sigma_{\A(r)}(x,\xi)$ of operator $\A(r)$ satisfies Ineq.~\eqref{Ineq_CV} of 
Section~\ref{Sec_CV} with upper bounds $C_{\bsalpha}$ that may be chosen 
independent of $r\ge1$. Consequently, the eigenvalues $\lambda_k^{(r)}$ of $\A(r)$ are 
uniformly bounded for $r\ge1$; define
\begin{equation} \label{Lambda}
  \Lambda_p\triangleq\max\left\{\sup_{\,r\ge1}\lambda_0^{(r)},\max_{x,\xi}|p(x,\xi)|^2\right\}.
\end{equation}
This constant appears in the next theorem:
\begin{theorem}[Szeg\H{o} Theorem]\label{SzegoTh} Let $g:[0,\Delta]\rightarrow\mathbb{R}$, 
$\Delta\in(0,\infty)$, be a continuous function such that $\lim_{x\rightarrow 0+}g(x)/x$ exists. For 
any functions $a,\,b:[1,\infty)\rightarrow\mathbb{R}$, where $a(x)$ is bounded and 
$\Lambda_p b(x)\in[0,\Delta]$, define the function 
$G(x,z)=a(x)g(b(x)z),\,(x,z)\in[1,\infty)\times[0,\Lambda_p]$. Then it holds
\begin{equation}
  \sum_{k=0}^\infty G(r,\lambda_k^{(r)})\doteq\frac{1}{2\pi}
          \iint_{\mathbb{R}^2}G(r,|p_r(x,\xi)|^2)\,\d x\,\d\xi. \label{Szego}
\end{equation}
\end{theorem}
\begin{IEEEproof}
The function $f(x)=g(x)/x,\,x\in(0,\Delta],$ has a continuous extension $F(x)$ onto 
the compact interval $[0,\Delta]$. By virtue of the Weierstrass approximation theorem, for 
any $m\in\mathbb{N}$ there exists a polynomial $F_{N_m-1}(x)$ of some degree $N_m-1$ such that 
$|F(x)-F_{N_m-1}(x)|\le\epsilon_m=\frac{1}{m}$ for all $x\in [0,\Delta]$. Consequently, the polynomial 
$g_{N_m}(x)=xF_{N_m-1}(x)$ of degree $N_m$ satisfies the inequality
\begin{equation}
  |g(x)-g_{N_m}(x)|\le \epsilon_m x,\,x\in[0,\Delta]. \label{WAS_ineq}
\end{equation}

Define the polynomial with variable coefficients $G_{N_m}(x,z)=a(x)\,g_{N_m}\!(b(x)z)$. We now 
show that
\begin{equation}
  r^{-2}\sum_{k=0}^\infty G_{N_m}(r,\lambda_k^{(r)})\rightarrow
        r^{-2}\sum_{k=0}^\infty G(r,\lambda_k^{(r)}) \label{first_arrow}
\end{equation}
and
\begin{eqnarray}
\lefteqn{\frac{r^{-2}}{2\pi}\iint G_{N_m}(r,|p_r(x,\xi)|^2)\,dx\,d\xi}\nonumber\\
        &\rightarrow&\frac{r^{-2}}{2\pi}\iint G(r,
	                             |p_r(x,\xi)|^2)\,dx\,d\xi \label{second_arrow}
\end{eqnarray}
as $m\rightarrow\infty$, uniformly for all $r\ge1$ . To this end, first observe that by Eq.~\eqref{ID1} 
(generalized to the operator $\P_r,r\ge1$) it holds
\begin{equation}  \label{ID2}
  \begin{split}
    \sum_{k=0}^\infty \lambda_k^{(r)}&=\frac{1}{2\pi}\iint|p_r(x,\xi)|^2\,\d x\,\d\xi \\
	                             &=c_pr^2,
  \end{split}
\end{equation}
where $c_p=(2\pi)^{-1}\iint|p(x,\xi)|^2\,\d x\,\d\xi$ is a finite constant.

\textit{Proof of (\ref{first_arrow}):} By Ineq. (\ref{WAS_ineq}) we get (precluding the trivial case 
$\Lambda_p=0$)
\begin{eqnarray*}
\lefteqn{|\sum_{k=0}^\infty G(r,\lambda_k^{(r)})-\sum_{k=0}^\infty G_{N_m}(r,\lambda_k^{(r)})|}\\
     &\le& \sum_{k=0}^\infty|G(r,\lambda_k^{(r)})-G_{N_m}(r,\lambda_k^{(r)})|\\
     &\le& M\epsilon_m(\Delta/\Lambda_p)\sum_{k=0}^\infty \lambda_k^{(r)},
\end{eqnarray*}
where $M=\sup_{x\ge1}|a(x)|<\infty$. Since $\sum_{k=0}^\infty\lambda_k^{(r)}=c_pr^2$, after 
division of the inequality by $r^2$, convergence in (\ref{first_arrow}) follows as claimed.

\textit{Proof of (\ref{second_arrow}):} Similarly,
\begin{eqnarray*}
\lefteqn{|\iint G(r,|p_r(x,\xi)|^2)\,dx\,d\xi}\\
     &&-\iint G_{N_m}(r,|p_r(x,\xi)|^2)\,dx\,d\xi|\\
     &\le& M\epsilon_m(\Delta/\Lambda_p)\iint|p_r(x,\xi)|^2\,dx\,d\xi.
\end{eqnarray*}
Since $(2\pi)^{-1}\iint|p_r(x,\xi)|^2\,dx\,d\xi=c_pr^2$, after division by $2\pi r^2$ we 
come to the same conclusion as before.

Finally, choose a (large) number $m\in\mathbb{N}$, so that the left-hand sides in \eqref{first_arrow}, 
\eqref{second_arrow} become arbitrarily close to their respective limits. Replace function $G$ in 
Eq.~\eqref{Szego} with the polynomial $G_{N_m}$. Then, by Lemma~\ref{Lemma_1} and the uniform convergence 
in \eqref{first_arrow}, \eqref{second_arrow} the theorem follows. 
\end{IEEEproof}
Note that Theorem~\ref{SzegoTh} applies to operator $\Atilde(r)$ without any changes.

\section{Waterfilling Theorem for the Capacity of Linear Time-Varying Channels} \label{Sec_V}	
\subsection{Waterfilling in the Time--Frequency Plane}
The function $N_r,r\ge1,$ occurring in the next theorem is defined by $N_r(t,\omega)=N_1(t/r,\omega/r)$ 
where
\begin{equation}
    N_1(t,\omega)=\frac{\theta^2}{2\pi}\,|p(t,\omega)|^{-2}, \label{N1}
\end{equation}
$p=\sigma_{\P}$ being the Weyl symbol of operator $\P$. Recall that $p\in\S(\mathbb{R}^2)$. 
$O(\cdot)$ denotes the standard Landau big-O symbol and $x^+$ denotes the positive part of 
$x\in\mathbb{R}$, $x^+=\max\{0,x\}$.
\begin{theorem}\label{WFT1} Assume that the average energy $S$ of the input signal depends on 
$r$ such that $S(r)=O(r^2)$ as $r\rightarrow\infty$. Then for the capacity (in nats per transmission) of 
the LTV channel \eqref{LTV_Ch} it holds
\begin{equation}
   C\doteq \frac{1}{2\pi}\iint_{\mathbb{R}^2}\frac{1}{2}
       \ln\left(1+\frac{(\nu-N_r(t,\omega))^+}{N_r(t,\omega)}\right)\,\d t\,\d\omega,   \label{C}
\end{equation}
where $\nu$ is chosen so that
\begin{equation}
   S\doteq\iint_{\mathbb{R}^2}(\nu-N_r(t,\omega))^+\,\d t\,\d\omega.   \label{S}
\end{equation}
\end{theorem}
\begin{IEEEproof}
The first part of the proof is accomplished by waterfilling on the noise variances 
\cite[Th.~7.5.1]{Gallager}. Let $\nu_k^2=\theta^2/\lambda_k^{(r)}(\mbox{put }\theta^2/0=\infty),\,k=0,1,\ldots,$ 
be the noise variance in the $k$th subchannel of the discretized LTV channel \eqref{LTV_discr}. We exclude the
trivial case $S=0$. The ``water level" $\sigma^2$ is then uniquely determined by the condition
\begin{equation}
  S = \sum_{k=0}^\infty (\sigma^2-\nu_k^2)^+=\sum_{k=0}^{K-1} (\sigma^2-\nu_k^2),\label{def_sigma}
\end{equation}
where $K=\max\{k\in\mathbb{N};\nu_{k-1}^2<\sigma^2\}$ 
is the number of subchannels in the resulting finite-dimensional vector Gaussian channel. The capacity 
$C$ of that vector channel is achieved when the components $X_k$ of the input vector 
$(X_0,\ldots,X_{K-1})$ are independent random variables $\sim \mathcal{N}(0,\sigma^2-\nu_k^2)$; then
\begin{equation}
  C=\sum_{k=0}^{K-1}\frac{1}{2}\ln\left(1+\frac{\sigma^2-\nu_k^2}{\nu_k^2}\right)
                                                                      \quad\mathrm{nats}. \label{C1}   
\end{equation}

In the second part of the proof we apply the above Szeg\H{o} theorem, Theorem~\ref{SzegoTh}. To start 
with, note that $\sigma^2$ is dependent on $r$ and that always $\sigma^2=\sigma^2(r)>0$. Additionally, 
suppose for the time being that the function $\sigma^2(r)$ is finitely upper bounded as 
$r\rightarrow\infty$. Define 
\begin{equation}
  \ln_+ x=\left\{\!\!\begin{array}{cl}\max\{0,\ln x\} & \mbox{if }x>0,\\
                                                    0 & \mbox{if }x=0.
                     \end{array}\right.                                        \label{ln+}
\end{equation}
By Eq.~\eqref{C1} we now have
\begin{align*}
  C&=\sum_{k=0}^\infty\frac{1}{2}\ln_+\left(\frac{\sigma^2(r)}{\theta^2}\lambda_k^{(r)}\right)\\
   &=\sum_{k=0}^\infty a(r)g(b(r)\lambda_k^{(r)}),
\end{align*}
where $a(r)=1$, $b(r)=\sigma^2(r)/\theta^2$, $g(x)=\frac{1}{2}\ln_+x,x\in[0,\Delta]$, and $\Delta$ is 
chosen so that $\Lambda_p b(r)\le\Delta<\infty$ when $r$ is large enough, $\Lambda_p$ being the 
constant \eqref{Lambda}. This choice is possible since $\sigma^2(r)$ remains bounded as 
$r\rightarrow\infty$; w.l.o.g., we assume $\Lambda_p b(r)\in[0,\Delta]$ for \emph{all} $r\ge1$. 
Then, by Theorem~\ref{SzegoTh} it follows that $C=C(r)$ satisfies
\begin{align}
    C&\doteq\frac{1}{2\pi}\iint
       \frac{1}{2}\ln_+\left(\frac{\sigma^2(r)}{\theta^2}\,|p_r(x,\xi)|^2
                                                                     \right)\,\d x\,\d\xi \nonumber\\
     &=\frac{1}{2\pi}\iint\frac{1}{2}
 \ln\!\left[1+\frac{\left(\frac{\sigma^2(r)}{2\pi}-N_r(t,\omega)\right)^+}{N_r(t,\omega)}
                                                                     \right]\!\d t\,\d\omega, \label{C2}
\end{align}
where $N_r(t,\omega)=\frac{\theta^2}{2\pi}\,|p_r(t,\omega)|^{-2}$. Next, rewrite Eq.~\eqref{def_sigma} as
\[
   S=\sum_{k=0}^\infty\sigma^2(r)\left(1
                            -\frac{1}{\frac{\sigma^2(r)}{\theta^2}\lambda_k^{(r)}}\right)^+.
\]
Put $a(r)=\sigma^2(r)$, $b(r)=\sigma^2(r)/\theta^2$ and define
\[
    g(x)=\left\{\!\!\begin{array}{cl}\left(1-\frac{1}{x}\right)^+ & \mbox{if }x>0,\\
                                                                0 & \mbox{if }x=0.
                     \end{array}\right.
\] 
Again, w.l.o.g., we may assume that $a(r)$ is bounded and $\Lambda_p b(r)\in[0,\Delta]$ for \emph{all} 
$r\ge1$ where $\Delta$ is chosen as above. Then, by Theorem~\ref{SzegoTh} it follows that
\begin{align}
   S&\doteq\frac{1}{2\pi}\iint\sigma^2(r)\left(1-
      \frac{1}{\frac{\sigma^2(r)}{\theta^2}\,|p_r(x,\xi)|^2}\right)^+\,\d x\,\d\xi \nonumber\\
    &=\iint \left(\frac{\sigma^2(r)}{2\pi}-N_r(t,\omega)\right)^+\,\d t\,\d\omega.           \label{S2}
\end{align}
Finally, replacement of $\frac{\sigma^2(r)}{2\pi}$ in Eqs.~\eqref{C2}, \eqref{S2} by parameter 
$\nu$ yields Eqs. \eqref{C}, \eqref{S}.

We complete the proof by a bootstrap argument: Take Eq.~\eqref{S} as a true equation and use \emph{it} for 
the definition of $\sigma^2(=2\pi\nu)$; after a substitution we obtain
\[
  \iint (\nu-N_1(t,\omega))^+\,\d t\,\d\omega=S(r)/r^2.
\]
Because of the growth condition imposed on $S$, $\nu=\nu(r)$ stays below a finite upper bound as
$r\rightarrow\infty$, and so does $\sigma^2(r)$. Consequently, the previous argument applies and the 
capacity $C$ is given by 
Eq.~\eqref{C}. Second, by reason of Theorem~\ref{SzegoTh}, it holds for the actual average input energy 
$S_{\mathrm{act}}(r)$ $=\sum_{k=0}^\infty(\sigma^2(r)-\nu_k^2)^+$ that $S_{\mathrm{act}}\doteq S$. Thus, 
the \emph{dotted} equation \eqref{S} applies anyway---even when $S$ is taken as $S_{\mathrm{act}}$.
\end{IEEEproof}
From the property $p\in\S(\mathbb{R}^2)$ it is easily deduced that, say,
\[
  N_1(t,\omega)\ge c_1(t^2+\omega^2),\,(t,\omega)\in\mathbb{R}^2,
\]
where $c_1$ is some positive constant depending on $p$; therefore, condition \eqref{S} certainly 
makes sense.

Note that the use of Landau symbols in Theorem~\ref{WFT1} does not mean that we need to pass to the 
limit (here, as $r\rightarrow\infty$). Rather, the dotted equations \eqref{C}, \eqref{S} may give 
useful approximations even when $r$ is finite (but large enough).
\begin{example} \label{Example_1}
Consider the HS operator $\P$ on $L^2(\mathbb{R})$ with the bivariate Gaussian function
\begin{equation} 
  p(t,\omega)=\e^{-\frac{1}{2}(\gamma^{-2}t^2+\gamma^2\omega^2)},\label{Eq_WS_Exp1}
\end{equation}
$\gamma>0$ fixed, as the Weyl symbol. Then $\P_r,r\ge1,$ has the Weyl symbol 
$p_r(t,\omega)=\exp[-(\gamma^{-2}t^2+\gamma^2\omega^2)/(2r^2)]$.  $\P_r$ is related to the operator 
$\P _\delta^{(\gamma)}$ of the so-called heat channel \cite{Ham2014} by 
the equation $\P_r=c\,\P _\delta^{(\gamma)}$, where $\delta=2\,\mathrm{arccoth}(2r^2)>0$ and $c=\cosh(\delta/2)$. 
$\P^{(\gamma)}_\delta$ has the diagonalization \cite{Daub}, \cite{Ham2004}, \cite{Ham2014}
\[
  (\boldsymbol{P}_\delta^{(\gamma)}f)(t)=
                        \sum_{k=0}^\infty\rho^{k+\frac{1}{2}}\langle f,f_k\rangle f_k(t),
\]
where $\rho=\e^{-\delta}$ and $f_k(t)=(D_\gamma H_k)(t)\triangleq\gamma^{-\frac{1}{2}}H_k(t/\gamma)$ is the  dilated 
$k$th Hermite function $H_k(t)$; the real-valued eigenfunctions $f_k,\,k=0,1,\ldots,$ form an orthonormal system in 
$L^2(\mathbb{R})$. Therefore, $\A(r)=\PrStarPr=\P_r^2$ has the eigenvalues 
$\lambda_k^{(r)}=c^2\rho^{2k+1},k=0,1,\ldots,$ so that the LTV channel \eqref{LTV_Ch} reduces 
to the discrete vector channel \eqref{LTV_discr} where the noise random variables $Z_k\sim\mathcal{N}(0,\nu_k^2)$ 
have the variances $\nu_k^2=(\theta/c)^2\rho^{-2k-1}$. Take the average input energy 
$S(r)=2\pi r^2\theta^2\,\SNR$, where $\SNR>0$ is the signal-to-noise ratio ($2\pi r^2\theta^2$ having the
interpretation of the average energy of the relevant noise). In Fig.~\ref{Figure_3}, capacity values labeled ``exact" 
have been computed numerically by waterfilling on the noise variances, as given in the proof of 
Theorem~\ref{WFT1}. Note that the results do not depend on $\theta^2$.

From Theorem~\ref{WFT1}, after computation of the double integrals and elimination of parameter $\nu$ 
we get the equation
\begin{equation}
  C\doteq\frac{r^2}{8}\left[\LambertW_0((4\pi\,\SNR -1)/\e)+1\right]^2, \label{Eq_Examp1}
\end{equation}
where $\LambertW_0$ is the principal branch of the Lambert W function determined by the 
conditions $\LambertW(x)\exp[\LambertW(x)]=x$ for all $x\in[-\e^{-1},\infty)$ and $\LambertW(0)=0$ 
\cite{CGHJK}, \cite{Ham2009}. In Fig.~\ref{Figure_3}, the approximate capacity 
\eqref{Eq_Examp1} is plotted as a function of $r$ (labeled ``waterfilling"). Surprisingly, the approximation is good even 
for spreading factors close to one.

\begin{figure}
\centering
\includegraphics[width=3.5in]{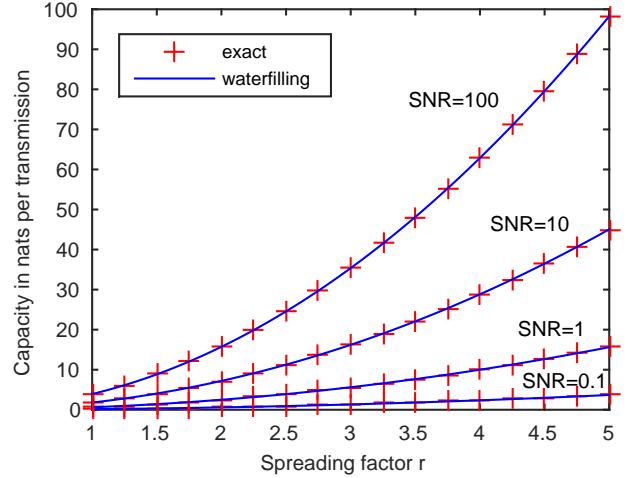}
\caption{Exact values and waterfilling approximation of the capacity of the LTV channel of 
Example~\ref{Example_1}}
\label{Figure_3}
\end{figure}
\end{example}

\subsection{Operational Meaning of the Capacity Result} \label{Sec_OperationalMeaning}
Theorem~\ref{WFT1} gives the \emph{information} capacity (in the sense of \cite{Cover}) of the LTV channel \eqref{LTV_Ch}. To provide 
this result with an operational meaning, we need to construct a code in the form of a set of continuous-time signals which achieves a rate arbitrarily 
close to this capacity along with constructive methods of encoding and decoding.

We use the notation in the proof of Theorem~\ref{WFT1}. For any fixed average input energy $S>0$ and any spreading factor $r\ge1$ held constant, 
the construction will be based on the eigenfunctions $f_k^{(r)}\!,\,k=0,\ldots,K-1$, of operator $\A(r)=\PrStarPr$, where $K$ is as in 
Eq.~\eqref{def_sigma}; at the receiver, the 
corresponding functions $g_k^{(r)}=\P_r f_k^{(r)}\!/[\lambda_k^{(r)}]^{\frac{1}{2}}$ will be used. Since the functions $g_k^{(r)}\!,f_k^{(r)}$ 
are in the range of the operators $\P_r,\P_r^*$ with Weyl symbols $p_r,\bar{p}_r\in\S(\mathbb{R}^2)$, resp., these functions are 
rapidely decreasing, $g_k^{(r)}\!,f_k^{(r)}\in\S(\mathbb{R})$. In practice, any finite collection of functions 
$u_1,\ldots,u_N\in\S(\mathbb{R})$ may be regarded to be concentrated on a common bounded interval centered at the origin and to be 
almost zero outside. Thus, for the sake of simplicity, we shall assume that $f_k^{(r)}(t)=g_k^{(r)}(t)=0,\,k=0,\ldots,K-1,$ if 
$|t|\ge d/2$ for some $d\in(0,\infty)$; $d$ will have the meaning of a delay later on. It will be convenient to switch from natural logarithms to 
logarithms to the base 2 and so from nats to bits. Then, the (information) capacity $C_k$ of the $k$th subchannel, $k=0,\ldots,K-1$, figuring in 
the sum on the right-hand side of Eq.~\eqref{C1}, reads 
\[
  C_k=\frac{1}{2}\log_2\left(1+\frac{\sigma^2-\nu_k^2}{\nu_k^2}\right)\quad\mathrm{bits}.
\]
We treat the $K$ subchannels as independent Gaussian channels with the noise variance $\nu_k^2$ each and follow the classical 
approach of Shannon \cite{Shannon1949}, \cite{Cover}: For the $k$th subchannel, for any rate $R_k$ with $0<R_k<C_k$ and any $\epsilon>0$ generate 
a codebook $\{\boldsymbol{a}_k(m)=(a_{k0}(m),\ldots,a_{k,L_k-1}(m));\,m=1,2,\ldots,M_k\triangleq 2^{\lfloor R_k L_k\rfloor}\}\subseteq\mathbb{R}^{L_k}$ 
with the property that 1) $a_{kl}(m),l=0,\ldots,L_k-1$, are realizations of i.i.d. random variables $\sim\mathcal{N}(0,\sigma^2-\nu_k^2)$ 
and 2) the probability of a maximum likelihood decoding error is smaller than $\epsilon$ for every transmitted codeword 
$\boldsymbol{a}_k(m),m=1,2,\ldots,M_k$. We may assume that $L_0=\ldots=L_{K-1}=L$. For every message 
$\boldsymbol{m}=(m_0,\ldots,m_{K-1})\in\{1,2,\ldots,M_0\}\times\ldots\times\{1,2,\ldots,M_{K-1}\}$ form the pulses
\[
  u_l(\boldsymbol{m},t-ld)=\sum_{k=0}^{K-1}a_{kl}(m_k)f_k^{(r)}(t-ld),\,l=0,\ldots,L-1,
\]
and take the pulse train
\begin{equation}
  u(\boldsymbol{m},t)=\sum_{l=0}^{L-1}u_l(\boldsymbol{m},t-ld) \label{Eq_umt}
\end{equation}
as input signal to the \emph{physical} channel. During transmission over that channel, each pulse $u_l(\boldsymbol{m},t-ld)$ undergoes a 
distortion modeled by the LTV filter \eqref{Op_Pr}, and results in the deformed pulse
\[
  v_l(\boldsymbol{m},t-ld)=\sum_{k=0}^{K-1}[\lambda_k^{(r)}]^{\frac{1}{2}}a_{kl}(m_k)g_k^{(r)}(t-ld).
\]
Thus, the output signal of the physical channel is
\[
  y(\boldsymbol{m},t)=\sum_{l=0}^{L-1}v_l(\boldsymbol{m},t-ld)+n(t),
\]
where $n(t)$ is a realization of white Gaussian noise as in the LTV channel model \eqref{LTV_Ch}. For any of the $K$ subchannels, pass the signal 
$y(\boldsymbol{m},t)$ through the matched filter with impulse response $h_k(t)$ as given in Section~\ref{Sec_III}; sample the matched filter output signal 
at time $ld,\,l=0,\ldots,L-1$. Since $y(\boldsymbol{m},t)=v_l(\boldsymbol{m},t-ld)+n(t)$ if $|t-ld|\le d/2$, we again obtain 
estimates $\hat{a}_{kl}(m_k)=a_{kl}(m_k)+z_{kl}$ for $a_{kl}(m_k)$, where $z_{kl}$ are realizations of independent Gaussian random variables 
$\sim\mathcal{N}(0,\nu_k^2)$. Maximum likelihood decoding of the perturbed codeword 
$\tilde{\boldsymbol{a}}_k(m_k)\triangleq(\hat{a}_{k0}(m_k),\ldots,\hat{a}_{k,L-1}(m_k))$ yields the correct codeword $\boldsymbol{a}_k(m_k)$ (thus, 
$m_k$) with a probability of error smaller than $\epsilon$. At the transmitter, choose the message $\boldsymbol{m}$ at random such that each component $m_k$ has 
probability $M_k^{-1}$ and is independent of the other components; convey $\boldsymbol{m}$ through a pulse train as described. Then---treating each of the 
$K$ subchannels separately---the total rate $R_{\mathrm{tot}}=\frac{1}{L}\sum_{k=0}^{K-1}\lfloor R_kL\rfloor$ (in bits per 
pulse) is attained with a total probability of a decoding error smaller than $K\epsilon$. When $L\rightarrow\infty$, Shannon's theory \cite{Shannon1949} 
ensures that $\epsilon$ can be made as small as we wish. Moreover, $R_{\mathrm{tot}}\rightarrow R\triangleq R_0+\ldots+R_{K-1}$ and, by the law of large 
numbers, the average input energy $\frac{1}{L}\sum_{l=0}^{L-1}\sum_{k=0}^{K-1}a_{kl}^2(m_k)$ tends to $\sum_{k=0}^{K-1}(\sigma^2-\nu_k^2)=S$ with probability 1. 
Finally, since the rate $R$ may be chosen arbitrarily close to the capacity $C=C_0+\ldots+C_{K-1}$ (at the expense of a larger length $L$ of the pulse train), 
the construction of the desired coding system is complete.
\begin{example} \label{Example_2}
Consider the LTV channel \eqref{LTV_Ch} with the operator $\P_r=c\,\P^{(\gamma)}_\delta$ of Example~\ref{Example_1}. The eigenfunctions of operator 
$\A(r)=\PrStarPr$ are the functions $f_k^{(r)}(t)=f_k(t)=(D_\gamma H_k)(t),\,k=0,1,\ldots,$ (here, not depending on $r$); the functions $g_k^{(r)}$ in the 
SVD~\eqref{SVD_r} of $\P_r$ coincide with $f_k^{(r)}$ for all $k$. Now, choose specifically $r=2,\,\gamma=1/10$ and take the 
average input energy $S=2\pi r^2\theta^2\,\SNR$ (as generally assumed in Example~\ref{Example_1}) with $\SNR=100$ and noise PSD $N_0/2=\theta^2=0.01$ (unit omitted). 
Waterfilling on the noise variances $\nu_k^2=(\theta/c)^2\rho^{-2k-1}(\rho=\e^{-\delta}),\,k=0,1,\ldots,$ as given in the proof of Theorem~\ref{WFT1}, yields the 
number of $K=11$ subchannels. In Fig.~\ref{Figure_4}(a), the first $K$ eigenfunctions $f_k^{(r)}(t)=(D_\gamma H_k)(t),\,k=0,\ldots,K-1,$ are displayed. 
The portion of an input pulse train plotted in Fig.~\ref{Figure_4}(b) has been computed according to Eq.~\eqref{Eq_umt} with the delay parameter 
$d=6a$, $a=\sqrt{2}r\gamma$, by numerical simulation of the involved random variables. Observe that there is no appreciable overlap of individual pulses. 
Each pulse transmits 22.6~bits ($=$15.7~nats, cf.~Fig.~\ref{Figure_3}) of information arbitrarily reliably [provided that the length of the pulse 
train(s) becomes larger and larger]. The meaning of parameter $a$ will be explained in Section~\ref{Sec_VII}.
\begin{figure}
\centering
\includegraphics[width=3.5in]{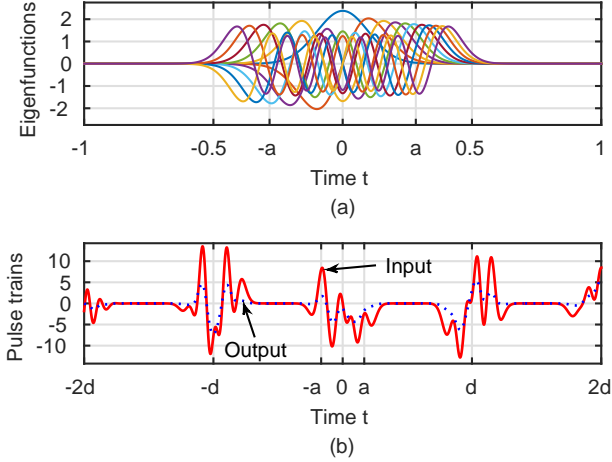}
\caption{(a) First eleven eigenfunctions, dilated Hermite functions, for the LTV channel of Example~\ref{Example_2}. (b) Portion of an input pulse 
train (centered at the origin) to the physical channel and the corresp. distorted output (without noise) of the same example. Time $t$ is measured in 
some unit of time; on the $y$-axis, also the physical dimension is omitted.}
\label{Figure_4}
\end{figure}
\end{example}

\subsection{Comparison with Classical Work}
Gallager's theorem \cite[Th.~8.5.1]{Gallager} gives the capacity of LTI channels under very 
general assumptions. In the case of an LTI filter with a bounded and square-integrable frequency response 
$H(\omega)=\int_{-\infty}^\infty \e^{-\i\omega t}h(t)\,\d t$ (a.k.a. transfer function; $h\not=0$ is the impulse 
response) and additive white Gaussian noise of PSD $N_0/2=\theta^2>0$ at the filter output, Gallager's theorem states that the 
capacity (in bits per second) is given parametrically by
\begin{align}
  C&= \frac{1}{2\pi}\int_{-\infty}^\infty\frac{1}{2}
                    \log_2\left(1+\frac{(\nu-N(\omega))^+}{N(\omega)}\right)\,\d\omega   \label{C_LTI}\\
  S&=\int_{-\infty}^\infty(\nu-N(\omega))^+\,\d\omega,	                                  \label{S_LTI}		  
\end{align}
where $\nu$ is the parameter, $S$ is average input \textit{power}, and
\begin{equation}
  N(\omega)=\frac{\theta^2}{2\pi}\,|H(\omega)|^{-2}.                            \label{N}
\end{equation}
We observe a perfect formal analogy between the waterfilling formulas \eqref{C_LTI}, \eqref{S_LTI} and those in 
Theorem~\ref{WFT1}. Moreover, the functions \eqref{N} and \eqref{N1} are the reciprocal 
squared modulus of the (time--frequency) transfer function of the respective filter times the same noise figure.

Eqs.~\eqref{C}, \eqref{S} may also be used, of course, for a parametric representation of the 
function $C(S)$ with $\nu$ as parameter.

\section{Reverse Waterfilling Theorem for Related Nonstationary Sources} \label{Sec_VI}
In the present section, we consider the nonstationary source formed by the nonstationary zero-mean 
Gaussian process given by the Karhunen--Lo\`{e}ve expansion
\begin{equation}
  X(t)=\sum_{k=0}^\infty X_k\,g_k^{(r)}(t),\,t\in\mathbb{R},\label{KL}
\end{equation}
where the coefficients $X_k,\,k=0,1,\ldots,$ are independent random variables 
$\sim\mathcal{N}(0,\sigma_k^2)$ with the variances 
$\sigma_k^2=\sigma^2\lambda_k^{(r)},\,\sigma>0$. This is the response of the LTV 
filter~\eqref{SVD_r} to white Gaussian noise with PSD $N_0/2=\sigma^2$; cf. \cite{Gallager}. 
This source is depicted in Fig.~\ref{Fig_2}. 
\subsection{Wigner--Ville Spectrum of the Source} \label{Sec_VI_A}
In the present subsection, the spreading factor $r\ge1$ is initially not essential, hence set to one and 
not displayed. 

The Wigner--Ville spectrum (WVS) $\Phi(t,\omega)$ of the nonstationary random process 
$\{X(t),t\in\mathbb{R}\}$ in \eqref{KL} describes its density of (mean) energy in the time--frequency 
plane \cite{FlandrinMartin}. The WVS may be regarded as the nonstationary counterpart to the PSD of a 
stationary random process. It is defined by means of the Wigner distribution $Wx$ of 
the realizations $x(t)$ of $\{X(t)\}$ and then taking the expectation \cite{FlandrinMartin}. Since 
$x(t)$ is almost surely in $L^2(\mathbb{R})$, we may write
\[ 
  (Wx)(t,\omega)=\frac{1}{2\pi}\int_{-\infty}^\infty \e^{-\i\omega t'}
        x\left(t+\frac{t'}{2}\right)\overline{x\left(t-\frac{t'}{2}\right)}\d t'.
\]
The WVS $\Phi(t,\omega)=\mathsf{E}[(WX)(t,\omega)]$ of the random process $\{X(t)\}$ is then 
given by
\begin{equation}
  \Phi(t,\omega)=\frac{1}{2\pi}\int_{-\infty}^{\infty}\e^{-\i\omega t'}
                       \mathscr{R}\left(t+\frac{t'}{2},t-\frac{t'}{2}\right)\d t',\label{W-Kh}
\end{equation}
where $\mathscr{R}(t_1,t_2)=\mathsf{E}[X(t_1)\overline{X(t_2)}]$ is the autocorrelation function. 
Appropriately enough, identities such as \eqref{W-Kh} are called a nonstationary Wiener--Khinchine 
theorem in \cite{Kozek96}. A computation yields
\[
  \mathscr{R}(t_1,t_2)=\sigma^2\sum_{k=0}^\infty\lambda_kg_k(t_1)\overline{g_k(t_2)}
  =\sigma^2K_{\Atilde}(t_1,t_2),
\]
where $K_{\Atilde}$ is the kernel of the operator $\Atilde=\P\P^*$, see Eq.~\eqref{K_PPstar}. By means of 
the Wigner transform \eqref{WT}, the Weyl symbol of $\Atilde$ becomes
\begin{equation}
  \sigma_{\Atilde}(t,\omega)=\int_{-\infty}^\infty \e^{-\i\omega t'}K_{\Atilde}\left(t+\frac{t'}{2},
                                               t-\frac{t'}{2}\right)\,\d t'. \label{Eq_sigmaAtilde}
\end{equation}
Comparing Eqs.~\eqref{Eq_sigmaAtilde} and \eqref{W-Kh} we thus obtain
\[
  \Phi(t,\omega)=\frac{\sigma^2}{2\pi}\cdot\sigma_{\Atilde}(t,\omega).
\]

In the general case $r\ge1$, the WVS depends on $r$ and we shall write $\Phi(r,\cdot,\cdot)$ for it; then, 
the latter equation becomes
\begin{equation}
  \Phi(r,t,\omega)=\frac{\sigma^2}{2\pi}\cdot\sigma_{\Atilde(r)}(t,\omega).   \label{WVS}
\end{equation}
By use of the trace rule \eqref{Eq_tracerule} and Eq.~\eqref{ID1} (rewritten for $\Atilde$ and then 
generalized to $r\ge1$) we conclude that
\begin{align*}
  \iint\Phi(r,t,\omega)\,\d t\,\d\omega
     &=\sigma^2\cdot\frac{1}{2\pi}\iint\sigma_{\Atilde(r)}(t,\omega)\,\d t\,\d \omega\\
     &=\sigma^2\cdot\tr\,\Atilde(r)\\
     &=\sum_{k=0}^\infty\sigma^2\lambda_k^{(r)},
\end{align*}
where the last infinite sum is indeed the average energy $E(r)=\sum_{k=0}^\infty\sigma_k^2$ of the 
realizations $x(t)$ of the random process \eqref{KL}; Eq.~\eqref{ID2} yields 
\begin{equation}
  E(r)=c_pr^2\sigma^2.  \label{E}
\end{equation}

By means of Lemma~\ref{power_lemma} we get from \eqref{WVS} for the WVS the asymptotic expansion
\[
  \Phi(r,t,\omega)\sim\frac{\sigma^2}{2\pi}\left(|p_r(t,\omega)|^2+
                                        \sum_{k=1}^\infty r^{-2k}\tilde{a}_k(t/r,\omega/r)\right),
\]
where $\tilde{a}_k\in\S(\mathbb{R}^2)$. The expression 
$\frac{\sigma^2}{2\pi}\,|p_r(t,\omega)|^2$---call it principal \textit{term} of the WVS 
$\Phi(r,t,\omega)$---will play a prominent role in the next subsection.
\begin{remark}
Asymptotically, the principal term might be a good substitute for the WVS $\Phi(r,t,\omega)$ itself. 
It is not only similar in shape, but it also gives the same average energy [see \eqref{ID2}] and is 
non-negative throughout (cf. \cite{Flandrin}).
\end{remark}

\subsection{Reverse Waterfilling in the Time--Frequency Plane} \label{Sec_VI_B}
Substitute the continuous-time Gaussian process $\{X(t),\,t\in\mathbb{R}\}$ in (\ref{KL}) by the
sequence of coefficient random variables $\boldsymbol{X}=X_0,X_1,\ldots\,$. For an estimate 
$\boldsymbol{\hat{X}}=\hat{X}_0,\hat{X}_1,\ldots$ of $\boldsymbol{X}$ we take the squared-error 
distortion $D=\mathsf{E}[\sum_{k=0}^\infty(X_k-\hat{X}_k)^2]$ as distortion measure. In our context, 
$D$ depends on $r$ and it always holds $0<D(r)\le E(r)$, where $E(r)$ is as in \eqref{E}.

\subsubsection{Computation of the rate distortion function} In the next theorem, the function 
$\Phi_r,r\ge1,$ is defined by $\Phi_r(t,\omega)=\Phi_1(t/r,\omega/r)$ where
\[
    \Phi_1(t,\omega)=\frac{\sigma^2}{2\pi}\,|p(t,\omega)|^2,
\]
$p\in\S(\mathbb{R}^2)$ being the Weyl symbol of operator $\P$. Recall that
\[
   \iint_{\mathbb{R}^2}\Phi_r(t,\omega)\,\d t\,\d\omega=E(r).
\]
The Landau symbol $\Omega(\cdot)$ is defined 
for any two functions as in Def.~\ref{def_2} as follows: $A(x)=\Omega(B(x))$ as $x\rightarrow\infty$ if 
$B(x)>0$ and $\liminf_{x\rightarrow\infty}A(x)/B(x)>0$.
\begin{theorem}\label{WFT2} Assume that the foregoing average distortion $D$ depends on 
$r$ such that $D(r)=\Omega(r^2)$ as $r\rightarrow\infty$. Then 
the rate distortion function $R = R(D)$ for the nonstationary source (\ref{KL}) is given by
\begin{equation}
  R\doteq\frac{1}{2\pi}\iint_{\mathbb{R}^2}\max\left\{0,\frac{1}{2}
             \ln\frac{\Phi_r(t,\omega)}{\lambda}\right\}\,\d t\,\d\omega,   \label{R}
\end{equation}
where $\lambda$ is chosen so that
\begin{equation}
   D\doteq\iint_{\mathbb{R}^2}
             \min\left\{\lambda,\Phi_r(t,\omega)\right\}\,\d t\,\d\omega.          \label{D}
\end{equation}
The rate is measured in nats per realization of the source.
\end{theorem}
\begin{IEEEproof}
The reverse waterfilling argument for a finite number of independent Gaussian sources \cite{Berger}, 
\cite{Cover} carries over to our situation without changes, resulting in a finite collection of Gaussian 
sources $X_0,\ldots,X_{K-1}$ where $K=\max\{k\in\mathbb{N};\sigma_{k-1}^2>\theta^2\}$ and the 
``water table" $\theta^2$ is chosen as the smallest positive number satisfying the condition
\begin{equation}
  D=\sum_{k=0}^\infty \min\{\theta^2,\sigma_k^2\}. \label{D_def}
\end{equation}
We exclude the trivial case $D=E(r)$. Then $K\ge1$ and the necessary rate $R=R(D)$ for the 
parallel Gaussian source $(X_0,\ldots,X_{K-1})$ amounts to \cite[Th.~10.3.3]{Cover}
\begin{equation}
  R = \sum_{k=0}^{K-1}\frac{1}{2}\ln\frac{\sigma_k^2}{\theta^2}\quad\mathrm{nats}. \label{R_def}
\end{equation}

Now we apply the above Szeg\H{o} theorem, Theorem~\ref{SzegoTh}. Again, $\theta^2$ depends on $r$. 
Suppose for the time being that $\theta^2=\theta^2(r)$ is finitely upper bounded for $r\ge1$ and positively 
lower bounded as $r\rightarrow\infty$. By Eq.~\eqref{D_def} we have
\begin{align*}
  D&=\sum_{k=0}^\infty\theta^2(r)\min\left\{1,\frac{\sigma^2}{\theta^2(r)}\lambda_k^{(r)}\right\}\\
   &=\sum_{k=0}^\infty a(r)g(b(r)\lambda_k^{(r)}),
\end{align*}
where $a(r)=\theta^2(r)$, $b(r)=\sigma^2/\,\theta^2(r)$, 
$g(x)=\min\{1,x\}$, $x\in[0,\Delta]$, and $\Delta$ is chosen so that $\Lambda_p b(r)\le\Delta<\infty$ when 
$r$ is large enough, $\Lambda_p$ being the constant \eqref{Lambda}. This choice is 
possible since $\theta^2(r)$ is positively lower bounded as $r\rightarrow\infty$; w.l.o.g., we assume here and 
hereafter that $\Lambda_p b(r)\in[0,\Delta]$ for \emph{all} $r\ge1$. Already, $a(r)$ is bounded for $r\ge1$. 
Then, from Theorem~\ref{SzegoTh} we infer that
\begin{align}
  D&\doteq\frac{1}{2\pi}\iint\theta^2(r)
    \min\left\{1,\frac{\sigma^2}{\theta^2(r)}\,|p_r(x,\xi)|^2\right\}\,\d x\,\d\xi        \nonumber \\
   &=\iint\min\left\{\frac{\theta^2(r)}{2\pi},\Phi_r(t,\omega)\right\}\,\d t\,\d\omega, \label{D2}
\end{align}
where $\Phi_r(t,\omega)=\frac{\sigma^2}{2\pi}\,|p_r(t,\omega)|^2$. Next, rewrite Eq.~\eqref{R_def} as
\[
  R=\sum_{k=0}^\infty\frac{1}{2}\ln_+\left(\frac{\sigma^2}{\theta^2(r)}\lambda_k^{(r)}\right),
\]
where $\ln_+$ is as defined in \eqref{ln+}. Taking $a(r)=1$, $b(r)=\sigma^2/\,\theta^2(r)$, 
$g(x)=\frac{1}{2}\ln_+x,x\in[0,\Delta]$, $\Delta$ chosen as before, by Theorem~\ref{SzegoTh} it follows 
that
\begin{align}
  R&\doteq\frac{1}{2\pi}\iint
       \frac{1}{2}\ln_+\left(\frac{\sigma^2}{\theta^2(r)}\,|p_r(x,\xi)|^2\right)\,
                                                                            \d x\,\d\xi \nonumber \\
   &=\frac{1}{2\pi}\iint\frac{1}{2}
       \ln_+\left[\frac{\Phi_r(t,\omega)}{\frac{\theta^2(r)}{2\pi}}\right]\,\d t\,\d\omega. \label{R2}
\end{align}
Finally, replacement of $\frac{\theta^2(r)}{2\pi}$ in Eqs.~\eqref{R2}, \eqref{D2} by the parameter 
$\lambda$ yields Eqs. \eqref{R}, \eqref{D}.
     
Again, we complete the proof by a bootstrap argument: Take Eq.~\eqref{D} as a true equation and use \emph{it} 
for the definition of $\theta^2(=2\pi\lambda)$; after a substitution we obtain
\[
  \iint\min\{\lambda,\Phi_1(t,\omega)\}\,\d t\,\d\omega=D(r)/r^2.
\]
Because of the growth condition imposed on $D$, $\lambda=\lambda(r)$ stays above a positive lower bound 
as $r\rightarrow\infty$ and so does $\theta^2(r)$. Moreover, always $\theta^2(r)\le2\pi\lambda_{\mathrm{max}}$ may 
be chosen where $\lambda_{\mathrm{max}}\triangleq\max_{t,\omega}\Phi_1(t,\omega)$. The rest of the argument 
follows along the same lines as in the proof of Theorem~\ref{WFT1}.
\end{IEEEproof}

\begin{example} \label{Example_3}
Consider the same ``Gaussian" LTV filter (operator) $\P$ with $\P_r=c\,\P^{(\gamma)}_\delta$ as in Example~\ref{Example_1}. 
The coefficients $X_0,X_1,\ldots$ of the random process $\{X(t)\}$ in \eqref{KL} then form a sequence of 
independent random variables $\sim\mathcal{N}(0,\sigma_k^2)$ with the variances 
$\sigma_k^2=(c\sigma)^2\rho^{2k+1}$ (cf. \cite{Ham2014}). For any average energy $E(r)=2^{-1}r^2\sigma^2$ of $\{X(t)\}$ 
define the distortion by $D(r)=E(r)/\SDR$, where the signal-to-distortion ratio $\SDR$ is at least one. In 
Fig.~\ref{Figure_5}, ``exact" rates $R$ have been computed numerically by reverse waterfilling on the 
signal variances, as given in the proof of Theorem~\ref{WFT2}.

From the two equations in Theorem~\ref{WFT2} we obtain by elimination of parameter $\lambda$ the 
closed-form equation
\begin{equation}
  R\doteq \frac{r^2}{8}\left[\LambertW_{-1}(-1/(\e\cdot\SDR))+1\right]^2,  \label{Eq_Examp2}
\end{equation}
where $\LambertW_{-1}$ is the branch of the Lambert W function determined by the 
conditions $\LambertW(x)\exp[\LambertW(x)]=x$ for all $x\in[-\e^{-1},0)$ and 
$\LambertW(x)\rightarrow -\infty$ as $x\rightarrow 0-$ \cite{CGHJK}, \cite{Ham2009}. In Fig.~\ref{Figure_5}, the 
approximate rate \eqref{Eq_Examp2} is plotted against $r$ (labeled ``reverse waterfilling"). Again, we observe a surprisingly 
good approximation even for spreading factors close to one.

\begin{figure}
\centering
\includegraphics[width=3.5in]{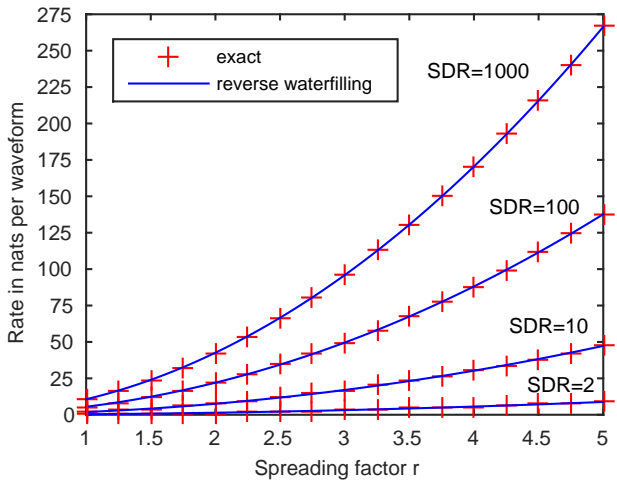}
\caption{Exact values and reverse waterfilling approximation of the rate for the nonstationary source 
of Example~\ref{Example_3}}
\label{Figure_5}
\end{figure}
\end{example}

\subsubsection{Comparison with classical work}
In Theorem~\ref{WFT2}, Eqs. \eqref{R}, \eqref{D} may also be used for a parametric representation of the rate distortion 
function $R(D)$. In parametric form, $R(D)$ has been given by Berger \cite{Berger} for a broad class of 
stationary random processes. In the latter parametric interpretation, Eq.~\eqref{R} is in perfect analogy 
to \cite[Eq.~(4.5.52)]{Berger} [with the (principal term of) WVS instead of the PSD], likewise Eq.~\eqref{D} with regard to 
\cite[Eq.~(4.5.51)]{Berger} (apart from a factor $\frac{1}{2\pi}$).

\section{A Lower Bound for the Spreading Factor} \label{Sec_VII}
Until now there has been no indication on how large the spreading factor $r$ should at least be chosen so that 
the dotted equations in the above waterfilling theorems yield useful approximations. The purpose of the 
present section is to identify a presumed lower bound for $r$.

For any $r\ge1$ define the operator $\Ahat(r):L^2(\mathbb{R})\rightarrow L^2(\mathbb{R})$ by the Weyl 
symbol $\sigma_{\Ahat(r)}(x,\xi)=2\pi\rho(r,x,\xi)$, where
\begin{equation}
  \rho(r,x,\xi)=\frac{\sigma_{\Atilde(r)}(x,\xi)}{\iint\sigma_{\Atilde(r)}(x',\xi')\,\d x'\,\d\xi'}\,.
                                                                                            \label{rho}
\end{equation}
Then $\Ahat(r)$ is self-adjoint, positive, of trace class with the trace 
$\tr\,\Ahat(r)=\iint \rho(r,x,\xi)\,\d x\,\d\xi=1$. Thus, $\Ahat(r)$ is a density 
operator and the Robertson--Schr\"{o}dinger uncertainty inequality (RSUI) applies \cite{Gosson11}; it 
reads: For any density operator on $L^2(\mathbb{R})$ with a Weyl symbol of the form $2\pi\rho(x,\xi)$, 
define the moments (for convenience, put $x_1=x$, $x_2=\xi$)
\begin{gather}
  \mu_i=\iint x_i\rho(x_1,x_2)\,\d x_1\,\d x_2, \label{mu}\\
  \sigma_{ij}=\iint (x_i-\mu_i)(x_j-\mu_j)\rho(x_1,x_2)\,\d x_1\,\d x_2     \label{sig}
\end{gather}
and write $\sigma_i^2=\sigma_{ii},\,i,j=1,2.$ Then it holds
\begin{equation} 
  \sigma_1^2\sigma_2^2\ge \sigma_{12}^2+\frac{\hbar^2}{4}, \label{RSUI}
\end{equation}
where $\hbar$ is the reduced Planck constant (which in our context is always set to one).

Now replace $\rho(x,\xi)$ with $\rho(r,x,\xi)$; since $\rho(r,x,\xi)$ depends on $r$, we shall 
write $\mu_i(r),\,\sigma_i^2(r),\,\sigma_{ij}(r)$ for its moments \eqref{mu}, \eqref{sig}. 
Although $\rho(r,x,\xi)$ is not a true probability density function (PDF), since it may assume negative 
values, its covariance matrix
\[
  \Sigma(r)=\left(\begin{array}{cc} \sigma_1^2(r)  & \sigma_{12}(r)\\
                                    \sigma_{12}(r) & \sigma_2^2(r)
                  \end{array}\right)
\]
is always positive definite (as is the covariance matrix of any density operator \cite{Narco}). The 
operators $\P_r,r\ge1,$ may also be viewed as time--frequency localization operators (TFLOs), comprising in part 
the TFLOs introduced by Daubechies \cite{Daub}.\footnote{Actually, the operator $\P^{(\gamma)}_\delta$ appearing in Example~\ref{Example_1} 
originates in such a TFLO (also called a Daubechies operator) with Gaussian weight in 
time and frequency; see \cite{Daub}, \cite{Ham2004}.} Since $\rho(r,t,\omega)$ is the normalized WVS $\Phi(r,t,\omega)$ discussed in 
Section~\ref{Sec_VI_A} [cf. Eq.~\eqref{WVS}], it is natural to define the ellipse of concentration (EoC) of $\P_r$ as 
the boundary of the region in phase space described by the inequality
\begin{equation}
  \big(x-\mu_1(r),\,\xi-\mu_2(r)\big)\,\Sigma(r)^{-1}\!
                                 \begin{pmatrix}x-\mu_1(r) \\ \xi-\mu_2(r)\end{pmatrix} \le4 \label{EoC}
\end{equation}
and having the property that the uniform distribution on it has the same first and second moments as the 
PDF at hand \cite{Cramer}. Since the EoC \eqref{EoC} has the area $A_{\mathrm{c}}=\pi\sqrt{\det(4\Sigma(r))}$, 
the RSUI can now be recast in the inequality 
$A_{\mathrm{c}}=4\pi\sqrt{\det\Sigma(r)}\ge4\pi\sqrt{\hbar^2/4}=2\pi$, or phrased in words: 
\emph{The area of the EoC of operator $\P_r,r\ge1,$ is at least $2\pi$.}

However, this is not a useful criterion since it holds for any $r$; to get a useful criterion, consider the 
(true) PDF
\begin{equation}
  \rho_r(x,\xi)\triangleq\frac{|p_r(x,\xi)|^2}{\iint|p_r(x',\xi')|^2\,\d x'\d\xi'},  \label{rho_pr}
\end{equation}
i.e., the normalized principal symbol of $\Atilde(r)$ [or $\A(r)$]. Note that the denominators in 
\eqref{rho} and \eqref{rho_pr} coincide,
\[
  \iint\sigma_{\Atilde(r)}(x,\xi)\,\d x\,\d\xi=\iint|p_r(x,\xi)|^2\,\d x\,\d\xi,
\]
which is a simple consequence of Eq.~\eqref{ID1} (in terms of $\Atilde$), Eq.~\eqref{Eq_tracerule} and a 
generalization to $r\ge1$; moreover, due to Lemma~\ref{power_lemma} it holds that
\begin{equation}
  \sigma_{\Atilde(r)}(x,\xi)=|p_r(x,\xi)|^2+r^{-2}R_1(r,x/r,\xi/r). \label{Eq_sigmaAtilde_r}
\end{equation}
The rationale is now as follows: When $r$ is large, $\rho_r(x,\xi)$ will be ``close to" $\rho(r,x,\xi)$; then the 
RSUI \eqref{RSUI} for $\rho(r,x,\xi)$ may be transposed to $\rho_r(x,\xi)$, resulting in a constraint on $r$. 
With this in mind, replace in \eqref{mu}, \eqref{sig} function $\rho(x_1,x_2)$ with $\rho_1(x_1,x_2)$ and denote 
the new values for $\mu_i,\,\sigma_i^2,\,\sigma_{ij}$ by $m_i,\,s_i^2,\,s_{ij}$, respectively. By means of 
Eq.~\eqref{Eq_sigmaAtilde_r} and observing that the common denominator in \eqref{rho}, \eqref{rho_pr} 
evaluates to $2\pi c_pr^2$, we then obtain $\mu_i(r)=m_ir+o(1)$ and by this $\sigma_{ij}(r)=s_{ij}r^2+o(r)$. 
Plugging the latter in the RSUI \eqref{RSUI} for $\rho(r,x,\xi)$ finally results in the desired constraint
\begin{equation}
  r^2\ge\frac{1}{2\sqrt{s_1^2s_2^2-s_{12}^2}}+o(1).  \label{criterion}
\end{equation}
Ineq.~\eqref{criterion} suggests a lower bound for the spreading factor $r$, thus providing the wanted criterion (in practice, 
the error term would be neglected). Note that asymptotically, i.e., as $r\rightarrow\infty$, Ineq.~\eqref{criterion} (with 
vanishing error term) becomes a necessary condition. 
\begin{example} \label{Example_4}
Consider the HS operator $\P$ on $L^2(\mathbb{R})$ with the Weyl symbol $p\in\S(\mathbb{R}^2)$ as given in 
Eq.~\eqref{Eq_WS_Exp1} of Example~\ref{Example_1} for any fixed parameter $\gamma>0$. Then the Weyl symbol 
$p_r$ of operator $\P_r,r\ge1,$ satisfies $\iint|p_r(x,\xi)|^2\,\d x\,\d\xi=\pi r^2$, so that the PDF \eqref{rho_pr} 
becomes
\begin{equation}
  \rho_r(x,\xi)=\frac{1}{\pi r^2}\,\e^{-\frac{1}{r^2}(\gamma^{-2}x^2+\gamma^2\xi^2)}. \label{Eq_rho_pr_HC}
\end{equation}
An evaluation of the integrals in \eqref{mu}, \eqref{sig} yields 
$m_1=m_2=0,$ $s_1^2=\gamma^2/2,\,s_2^2=\gamma^{-2}/2 $ and $s_{12}=s_{21}=0$. 
Consequently, Ineq.~\eqref{criterion} turns into
\[
   r^2\ge1+o(1),
\]
which, neglecting the error term, means no restriction at all. In fact, in Fig.~\ref{Figure_3} and 
Fig.~\ref{Figure_5} the approximation is already acceptable for spreading factors close to one. 

Finally, we add the explanation of the parameter $a$ occurring in Example~\ref{Example_2} of Section~\ref{Sec_OperationalMeaning}. 
To this end, we determine the EoC \eqref{EoC} of the above operator $\P_r,\,r\ge1,$ by the use of the identity 
$\P_r=c\,\P^{(\gamma)}_\delta$ (see Example~\ref{Example_1}). The Weyl symbol of the operator 
$\P^{(\gamma)}_\delta\!\circ(\P^{(\gamma)}_\delta)^*=\P^{(\gamma)}_{2\delta}$ is given in closed form in \cite{Ham2014}. By 
this means, Eq.~\eqref{rho} readily becomes
\[
  \rho(r,x,\xi)=\frac{1}{\pi\alpha\beta}\,\exp\left(-\frac{x^2}{\alpha^2}-\frac{\xi^2}{\beta^2}\right),
\]
where $\alpha=\gamma\sqrt{\coth\delta},\,\beta=\gamma^{-1}\sqrt{\coth\delta}$. The \emph{exact} EoC of the operator $\P_r$ 
is therefore the ellipse in phase space with the semi-axes $a_{\mathrm{x}}=\sqrt{2}\alpha,\,b_{\mathrm{x}}=\sqrt{2}\beta$ 
and the equation
\[
  x^2/a_{\mathrm{x}}^2+\xi^2/b_{\mathrm{x}}^2=1.
\]
From the PDF~\eqref{Eq_rho_pr_HC}, we obtain asymptotically, i.e., as $r\rightarrow\infty$, the approximate EoC with semi-axes 
$a=\sqrt{2}r\gamma,\,b=\sqrt{2}r/\gamma$. For instance, in the case of $r=2,\gamma=1/10$ we find the rather good approximations 
$a=0.2828,\,b=28.28$ (units omitted) of the exact values $a_{\mathrm{x}}=0.2850,\,b_{\mathrm{x}}=28.50$ (which is 
somewhat surprising since $r=2$ is still small). In Example~\ref{Example_2}, the foregoing value of $a$ has been used as an estimate of 
the effective half duration of a pulse.
\end{example}

\section{Conclusion}
Waterfilling theorems in the time--frequency plane for the capacity of an LTV 
channel with an average energy constraint and the rate distortion function for a related nonstationary source 
with a squared-error distortion constraint have been stated and rigorous 
proofs have been given. The waterfilling theorem for the LTV channel has been formulated in terms of the reciprocal squared modulus 
of the spread Weyl symbol of the LTV filter (times a noise figure), whereas in the reverse waterfilling 
theorem for the nonstationary source simply the squared modulus of the spread Weyl symbol (times a 
signal figure) has been used. The latter expression has been related to the WVS of the nonstationary source and 
recognized as its principal term. 
The LTV filter, initially an arbitrary HS operator, was later restricted to an operator
with a Weyl symbol in the Schwartz space of rapidly decreasing functions. This smoothness 
assumption was a prerequisite 
for a Szeg\H{o} theorem upon which the proofs of both waterfilling theorems rested in an 
essential way. A self-contained proof of the Szeg\H{o} theorem has been given. 
The formulas in the waterfilling theorems depend on the spreading factor and are asymptotic in nature. 
Two examples with a bivariate Gaussian function as the Weyl symbol showed that the waterfilling theorems 
may perform well even when the spreading factor is close to one. For the general case, based on an uncertainty inequality, a lower 
bound for the spreading factor has been suggested.

\appendix[Proof of Lemma \ref{power_lemma}]
In this appendix, we shall write $\boldsymbol{x}=(x_1,x_2)$ etc.\ for phase space points 
$(x,\xi)\in\mathbb{R}^2$ and $\d\x=\d x_1\d x_2$ etc.\ for the corresponding differential. Also, we use the 
notations $\langle\x\rangle\triangleq(1+x_1^2+x_2^2)^{1/2}$, 
$\Diamond_{\x}\triangleq1-\partial_{x_1}^2-\partial_{x_2}^2$ and write 
$\partial_{\x}^{\bsalpha}=\partial_{x_1}^{\alpha_1}\partial_{x_2}^{\alpha_2}$, 
${\x}^{\bsbeta}=x_1^{\beta_1}x_2^{\beta_2}$ with the multi-indices 
$\bsalpha=(\alpha_1,\alpha_2)$, $\bsbeta=(\beta_1,\beta_2)\in\mathbb{N}_0^2$. 
The following proof draws on \cite{Folland}, \cite{Oldfield}.

For any two operators $\P,\,\Q:L^2(\mathbb{R})\rightarrow L^2(\mathbb{R})$ with the Weyl symbols 
$p,\,q\in\S(\mathbb{R}^2)$, resp., the Weyl symbol of the product $\P\Q$, denoted by $p\,\#\,q$, is 
given by \cite{Folland}
\begin{multline}
  (p\,\#\,q)(\z)\\
     =\frac{1}{\pi^2}\int_{\mathbb{R}^2}\int_{\mathbb{R}^2} p(\z+\x)q(\z+\y)
                                                   \e^{2\i\det(\x;\y)}\,\d\x\,\d\y,  \label{Eq_hash}
\end{multline}
where $\det(\x;\y)= x_1y_2-x_2y_1$. Since we need to compute the Weyl symbol 
$p_r\#\,q_r,\,r\ge 1,$ we change to the more convenient operation $p\,\#_rq$ defined by 
$(p_r\#\,q_r)(\z)=(p\,\#_rq)(\z/r)$. A computation yields (see \cite{Folland} and \cite{Oldfield} in 
the case of $r=1$)
\[
  (p\,\#_rq)(\z)=\sum_{k=0}^{m-1} r^{-2k}a_k(\z)+r^{-2m}R_m(r,\z)
\]
with the functions $a_k\in\S(\mathbb{R}^2)$ given by $a_k(\z)=F_k(\z,\z)$, and
\begin{multline} \label{Eq_Rmrz}
  R_m(r,\z)=m\int_0^1(1-t)^{m-1}
              \bigg\{\frac{1}{\pi^2}\int_{\mathbb{R}^2}\int_{\mathbb{R}^2}\e^{2\i\det(\x;\y)}\\
                                                 \cdot F_m(\z+t\x/r,\z+\y/r)\,\d\x\,\d\y\bigg\}\d t,
\end{multline}
where
\begin{equation}
  F_k(\x,\y)=\frac{\i^k}{k!2^k}(\partial_{x_1}\partial_{y_2}-
                                      \partial_{x_2}\partial_{y_1})^k\,[p(\x)q(\y)]    \label{Eq_Fkxy}
\end{equation}
for $k=0,\ldots,m$. Note that $a_0(\z)=p(\z)q(\z)$.

First, we show that when $a\in\S(\cdot,\mathbb{R}^2),\,b\in\S(\mathbb{R}^2)$, then 
$c=a\,\#_rb\in\S(\cdot,\mathbb{R}^2),\,r\ge1$. To this end, note that for any positive integers $L,\,M$ 
it holds $\Diamond_{\y}^M\e^{2\i\det(\x;\y)}=\langle 2\x \rangle^{2M}\e^{2\i\det(\x;\y)}$ and 
$\Diamond_{\x}^L\e^{2\i\det(\x;\y)}=\langle 2\y \rangle^{2L}\e^{2\i\det(\x;\y)}$. By partial integration 
we then obtain from \eqref{Eq_hash} the representation
\begin{multline} \label{Eq_crz}
  c(r,\z)=\frac{1}{\pi^2}\int_{\mathbb{R}^2}\int_{\mathbb{R}^2}\e^{2\i\det(\x;\y)}
                           \Diamond_{\x}^L\frac{a(r,\z+\x/r)}{\langle2\x\rangle^{2M}}\\
                       \cdot\frac{\Diamond_{\y}^M[b(\z+\y/r)]}{\langle2\y\rangle^{2L}}\,\d\x\,\d\y.
\end{multline}
Concerning the computation of $\Diamond_{\x}^L(\cdot)$ occurring in \eqref{Eq_crz}, note that for any 
$\boldsymbol{\gamma}=(\gamma_1,\gamma_2)\in\mathbb{N}_0^2,\,0\le\gamma_1+\gamma_2\le 2L,$ it holds
\[
   |\partial_{\x}^{\boldsymbol{\gamma}}[\langle 2\x\rangle^{-2M}]|\le 
                                                      C\langle 2\x\rangle^{-2M},\x\in\mathbb{R}^2,
\]
where $C=C(M,L)<\infty$. Consequently, for any $\bsalpha,\bsbeta\in\mathbb{N}_0^2$, 
the expression $|\z^{\bsbeta}\partial_{\z}^{\bsalpha}c(r,\z)|$ may be upper bounded for all 
$\z\in\mathbb{R}^2$ and $r\ge1$ by a linear combination (with positive coefficients effectively not 
depending on $r$ since $1/r\le1$) of terms of the form
\begin{equation} \label{Eq_ii}
  \int_{\mathbb{R}^2}\int_{\mathbb{R}^2}\frac{|a_\lambda(r,\z+\x/r)b_\mu(\z+\y/r)|}
                                      {\langle2\x\rangle^{2M}\langle2\y\rangle^{2L'}}\,\d\x\,\d\y,
\end{equation}
where $a_\lambda\in\S(\cdot,\mathbb{R}^2)$, $b_\mu\in\S(\mathbb{R}^2)$ and $M\ge2$, $2\le L'\le L$ 
($L$ sufficiently large). Here we have used, possibly repeatedly, the fact that when 
$b_\mu\in\S(\mathbb{R}^2),$ then $z_ib_\mu(\z+\y/r)=\tilde{b}_\mu(\z+\y/r)-(y_i/r)b_\mu(\z+\y/r)$ where 
$\tilde{b}_\mu$ defined by $\tilde{b}_\mu(\z)=z_ib_\mu(\z)$ is again in $\S(\mathbb{R}^2)$. Replace 
the numerator of the integrand in \eqref{Eq_ii} by a constant upper bound and integrate. Summing up, we 
obtain the inequality
\begin{equation}
  |\z^{\bsbeta}\partial_{\z}^{\bsalpha}c(r,\z)|\le C_{\bsalpha\bsbeta}
                                                     <\infty,\,\z\in\mathbb{R}^2, \label{Eq_ineq_crz}
\end{equation}
where the constant $C_{\bsalpha\bsbeta}$ does not depend on $r\ge1$.

Second, we show that $R_m\in\S(\cdot,\mathbb{R}^2)$. The integral $I(t,r,\z)$ between braces 
$\{\ldots\}$ in \eqref{Eq_Rmrz} is a linear combination of expressions on the right-hand side of 
Eq.~\eqref{Eq_crz} after the substitution $a(r,\z+\x/r)\leftarrow(\partial_{\x}^{\bsalpha}p)(\z+t\x/r)$ 
and $b(\z+\x/r)\leftarrow(\partial_{\x}^{\bsbeta}q)(\z+\x/r)$, the partial derivatives (of order $m$) 
coming from those in \eqref{Eq_Fkxy}. Now let $\alpha,\beta\in\mathbb{N}_0^2$ be 
arbitrary. By the same reasoning as before, we infer that 
$|\z^{\bsbeta}\partial_{\z}^{\bsalpha}I(t,r,\z)|$ may be upper bounded by a linear combination (with 
positive coefficients effectively not depending on $r$ and $t$ since $t/r\le1$) of terms analogous to 
\eqref{Eq_ii}. Taking the supremum of the numerators of the integrands, we get rid of the variable $t$ 
so that the integral with respect to $t$ occurring in \eqref{Eq_Rmrz} may be computed (evaluating to 1). 
Again summing up, we obtain the analog to Ineq.~\eqref{Eq_ineq_crz}, where $c(r,\z)$ is to be replaced 
with $R_m(r,\z)$.

Now we are in a position to prove Eq.~\eqref{AE}. Take any function $b\in\S(\mathbb{R}^2)$. Then 
\begin{multline*}
  ((p\,\#_rq)\,\#_rb)(\z)=\sum_{k=0}^{m-1} r^{-2k}(a_k\,\#_rb)(\z)\\
                                                      +r^{-2m}R_{m,0}(r,\z),
\end{multline*}
where $R_{m,0}=R_m\,\#_rb\in\S(\cdot,\mathbb{R}^2)$ and
\[
  (a_k\,\#_rb)(\z)=\sum_{j=0}^{m-k-1} r^{-2j}a_{kj}(\z)+r^{-2(m-k)}R_{k,m-k}(r,\z)
\]
with the functions $a_{kj}\in\S(\mathbb{R}^2)$ and $R_{k,m-k}\in\S(\cdot,\mathbb{R}^2),$ 
$k=0,\ldots,m-1$. One readily finds
\[
  ((p\,\#_rq)\,\#_rb)(\z)=\sum_{k=0}^{m-1}r^{-2k}\tilde{a}_k(\z)+r^{-2m}\tilde{R}_m(r,\z),
\]
where the functions $\tilde{a}_k\in\S(\mathbb{R}^2)$ and $\tilde{R}_m\in\S(\cdot,\mathbb{R}^2)$ are given by
\[
  \tilde{a}_k(\z)=\sum_{i+j=k \atop i,j\ge0}a_{ij}(\z),\,\tilde{R}_m(r,\z)
                                                               =\sum_{k=0}^mR_{k,m-k}(r,\z).
\]
Note that $\tilde{a}_0=pqb$. Putting $q=\bar{p}$ and alternately $b=p$ or
$b=\bar{p}$ we obtain by induction, observing that brackets may be omitted, for the Weyl symbol 
$\sigma_{\A^n(r)}(\z)$ of $\A^n(r)$,
\[
  \underbrace{(p_r\#\,\bar{p}_r\#\ldots\#\,p_r\#\,\bar{p}_r)}_{\displaystyle n\;\mbox{factors}\;
                                                                                p_r\#\,\bar{p}_r}(\z)
                  =(p\,\#_r\bar{p}\,\#_r\ldots\#_rp\,\#_r\bar{p})(\z/r),
\]
the asymptotic expansion as given in Eq.~\eqref{AE} [written without superscripts $\tilde{}$ again and 
after the substitution $\z\leftarrow(x,\xi)$].

\section*{Acknowledgment}
The author wishes to thank the reviewers and the Associate Editor Prof. Daniela Tuninetti for their helpful 
comments, remarks, and suggestions.

\end{document}